\documentclass[useAMS,twocolumn,usenatbib]{mn2e}

\usepackage{graphicx}

\usepackage[fleqn]{amsmath}
\usepackage{amsfonts}
\usepackage{amssymb}

\usepackage[hyperindex]{hyperref}
\hypersetup{
breaklinks = {false},
colorlinks = {false},
linkcolor={black},
pdfpagemode = {None}, 
pdfborder = {0 0 1},
pdftitle = {Reducing distance errors for standard candles and standard sirens with weak-lensing shear and flexion maps},
pdfsubject = {},
pdfauthor = {S. Hilbert, J. R. Gair, L. J. King},
pdfkeywords = {gravitational lensing: weak, distance scale, gravitational waves, super novae: general, gamma-ray burst: general, cosmological parameters, cosmology: observations, cosmology: theory, methods: numerical}
}

\newcommand{\vect}[1]{\boldsymbol{#1}}
\newcommand{\tens}[1]{\boldsymbol{\mathsf{#1}}}

\newcommand{\ii}{\mathrm{i}}

\newcommand{\R}{\mathbb{R}}

\newcommand{\parder}[3][]{\frac{\partial^{#1} {#2}}{\partial {#3}^{#1}}}

\newcommand{\diff}[2][]{\mathrm{d}^{#1}{#2}}

\newcommand{\tr}{\ensuremath{\operatorname{tr}}}

\newcommand{\EV}[1]{\left\langle{#1}\right\rangle}
\newcommand{\bEV}[1]{\bigl\langle{#1}\bigr\rangle}

\newcommand{\kpc}{\ensuremath{\mathrm{kpc}}}
\newcommand{\Mpc}{\ensuremath{\mathrm{Mpc}}}
\newcommand{\Msolar}{\ensuremath{\mathrm{M}_\odot}}

\newcommand{\arcsect}{\ensuremath{\mathrm{arcsec}}}
\newcommand{\arcmint}{\ensuremath{\mathrm{arcmin}}}
\newcommand{\degt}{\ensuremath{\mathrm{deg}}}

\newcommand{\F}{\mathcal{F}}
\newcommand{\G}{\mathcal{G}}

\newcommand{\nablap}{{\ensuremath{\nabla_{\theta}^{+}}}}
\newcommand{\nablam}{{\ensuremath{\nabla_{\theta}^{-}}}}
\newcommand{\nablapm}{{\ensuremath{\nabla_{\theta}^{\pm}}}}

\newcommand{\zS}{z_\mathrm{S}}

\newcommand{\Dlum}{D_\text{lum}}

\newcommand{\ngal}{n_{\text{gal}}}
\newcommand{\pgal}{p_{\text{gal}}}
\newcommand{\zgal}{z_{\text{gal}}}
\newcommand{\thetagal}{\vect{\theta}_{\text{gal}}}

\newcommand{\gammagal}{\gamma_{\text{gal}}}
\newcommand{\Fgal}{\F_{\text{gal}}}
\newcommand{\Ggal}{\G_{\text{gal}}}

\newcommand{\sigmagammagal}{\sigma_{\gammagal}}
\newcommand{\sigmaFgal}{\sigma_{\Fgal}}
\newcommand{\sigmaGgal}{\sigma_{\Ggal}}

\newcommand{\thetagali}{\vect{\theta}^{(i)}_{\text{gal}}}
\newcommand{\wgali}{w^{(i)}_{\text{gal}}}

\newcommand{\gammagali}{\gamma^{(i)}_{\text{gal}}}

\newcommand{\kappaeff}{\kappa_{\text{eff}}}
\newcommand{\zmedian}{z_{\text{median}}}

\newcommand{\kappaest}{\kappa_{\text{est}}}
\newcommand{\hatkappaest}{\hat{\kappa}_{\text{est}}}
\newcommand{\kapparaw}{\kappa_{\text{raw}}}
\newcommand{\hatkapparaw}{\hat{\kappa}_{\text{raw}}}
\newcommand{\muest}{\mu_{\text{est}}}
\newcommand{\lambdaest}{\lambda_{\text{est}}}
\newcommand{\mures}{\mu_{\text{res}}}
\newcommand{\lambdares}{\lambda_{\text{res}}}

\newcommand{\kappaesti}{\kappa_{\text{est},i}}

\newcommand{\tlambdaest}{\tilde{\lambda}_{\text{est}}}
\newcommand{\lambdaresi}{\lambda_{\text{res},i}}

\newcommand{\thetas}{\theta_{\text{s}}}
\newcommand{\thetaw}{\theta_{\text{w}}}

\def\aj{AJ}%
\def\apj{ApJ}%
\def\apjl{ApJ}%
\def\apjs{ApJS}%
\def\aap{A\&A}%
\def\mnras{MNRAS}%
\def\prd{Phys.~Rev.~D}%
\def\nat{Nature}%
\def\physrep{Phys.~Rep.}%
\begin{document}

\title[Reducing distance errors with weak-lensing maps]{Reducing distance errors for standard candles and standard sirens with weak-lensing shear and flexion maps}
\author[S. Hilbert, J.R. Gair, \& L.J. King]{
Stefan Hilbert$^{1,2}$\thanks{\texttt{shilbert@astro.uni-bonn.de}},
Jonathan R. Gair$^{3}$, and 
Lindsay J. King$^{3,4}$
\\$^{1}$Argelander-Institut f{\"u}r Astronomie, Auf dem H{\"u}gel 71, 53121 Bonn, Germany
\\$^{2}$Max-Planck-Institut f{\"u}r Astrophysik, Karl-Schwarzschild-Stra{\ss}e 1, 85741 Garching, Germany
\\$^{3}$Institute of Astronomy, Madingley Road, CB3 0HA Cambridge, United Kingdom
\\$^{4}$Kavli Institute for Cosmology, Madingley Road, CB3 0HA Cambridge, United Kingdom
}
\date{\today}
\maketitle

\begin{abstract}
Gravitational lensing induces significant errors in the measured distances to high-redshift standard candles and standard sirens such as type-Ia supernovae, gamma-ray bursts, and merging supermassive black hole binaries. There will therefore be a significant benefit from correcting for the lensing error by using independent and accurate estimates of the lensing magnification.
Here we investigate how accurately the magnification can be inferred from convergence maps reconstructed from galaxy shear and flexion data.
We employ ray-tracing through the Millennium Simulation to simulate lensing observations in large fields, and perform a weak-lensing reconstruction on the simulated fields. We identify optimal ways to filter the reconstructed convergence maps and to convert them to magnification maps, and analyse the resulting relation between the estimated and true magnification for sources at redshifts $\zS = 1$ to 5.
We find that a deep shear survey with 100 galaxies/arcmin$^2$ can help to reduce the lensing-induced distance errors for standard candles/sirens at redshifts $\zS\approx1.5$ ($\zS\approx5$) on average by $20\%$ ($10\%$), whereas a futuristic survey with shear and flexion estimates from 500 galaxies/arcmin$^2$ yields much larger reductions of 50\% (35\%). For redshifts $\zS\geq3$, a further improvement by $\sim5\%$ can be achieved, if the individual redshifts of the galaxies are used in the reconstruction. Moreover, the reconstruction allows one to identify regions for which the convergence is low, and in which an error reduction by up to $75\%$ can be achieved.
Such strongly reduced magnification uncertainties will greatly improve the value of high-redshift standard candles/sirens as cosmological probes.
\end{abstract}

\begin{keywords}
gravitational lensing -- distance scale -- gravitational waves -- super novae: general -- gamma-ray burst: general
\end{keywords}

\section{Introduction}
\label{sec:Introduction}

As suggested by \citet{Lemaitre1927} and confirmed by \citet{Hubble1929}, there is a correlation between the distances and redshifts of galaxies that arises from the expansion of our Universe: galaxies at greater apparent distance show more redshifted spectral features. The exact relationship between distance and redshift changes for different cosmological theories or different cosmological parameters. Accurate measurements of the distance-redshift relation may therefore help to discriminate between models and to constrain cosmological parameters \citep{Linder2008_review}.

Distance estimates based on the apparent magnitudes of type-Ia supernova (SN) have provided constraints for the current expansion rate and the mean matter density of our Universe, and substantiated the evidence for the currently favoured $\Lambda$CDM model \citep[e.g.][]{KomatsuEtal2009}, since SN observations indicated a cosmological constant or another dark-energy component is currently accelerating the cosmic expansion \citep{RiessEtal1998,PerlmutterEtal1999}. In conjunction with other cosmological probes, current SN data even puts constraints on the properties of the dark energy \citep{KowalskiEtal2008,RubinEtal2009}.

While type-Ia SN can only be seen out to redshifts $z\sim2$, gamma-ray bursts (GRB) can be detected at much higher redshifts. There is growing evidence for correlations between observed and intrinsic burst properties \citep[e.g.][]{GhirlandaGhiselliniFirmani2006}, with which GRBs can provide distance estimates out to redshifts $z>6$ \citep{Schaefer2007}. When added to distance data at lower redshift, the higher-redshift bursts help to break parameter degeneracies in cosmological models and make GRBs a valuable discriminant between dark-energy models \citep{Schaefer2007,LiangEtal2008,AmatiEtal2008}.

Another promising way to obtain very accurate distance measurements in the future are observations of gravitational waves emitted by merging supermassive black hole binaries (MSBs) \citep{Schutz86,HolzHughes2005}. From the gravitational wave phasing, it is possible to determine the luminosity distance to the source and the redshifted mass, $M(1+z)$, to high precision. However, the redshift cannot be measured from the gravitational waves alone, so the use of MSBs as distance probes will rely on the identification of an electromagnetic counterpart to measure the redshift of the merger event \citep[][]{LangHughes2008}. Detection of the gravitational-wave signals from MSBs is one of the main goals of the planned Laser Interferometer Space Antenna\footnote{\texttt{http://lisa.nasa.gov/}} (LISA). MSBs are very luminous and can be observed out to the cosmic horizon. Current estimates suggest that LISA will see a few tens of events per year, of which most will have distance measurements to less than $\sim10\%$, and of the order of two per year will have distance measurements to less than $1\%$ and sky localisation to within a square degree \citep{LISApet}.

The distance measurements for standard candles like type-Ia SNe, GRBs, and standard sirens like MSBs are based on the comparison of the electromagnetic or gravitational radiation energy emitted by the source (assumed to be known at least statistically) and the radiation flux or amplitude received by the observer's detector. Apart from the emitted signal strength and the distance between source and observer (and other factors), the received signal strength is also affected by gravitational lensing: Gravitational deflection by intervening matter structures may focus or defocus the radiation towards the observer, resulting in an increased or decreased detector signal. This gravitational \mbox{(de-)}magnification induces an additional scatter in the distance estimates \citep[see, e.g.,][ and references therein]{HolzLinder2005}.

Evidence for gravitational \mbox{(de-)}magnification has been found in the GOODS SN data \citep{JoenssonEtal2006} and the Supernova Legacy Survey \citep{JoenssonEtal2010}. The lensing effect is small compared to other sources of uncertainty for current SN samples \citep[e.g.][]{RiessEtal2004}, but it will become significant for large future high-redshift (i.e. with $z\gtrsim1.5$) SN samples \citep{HolzLinder2005,JoenssonEtal2010}. In particular, pencil-beam surveys might be compromised by gravitational lensing because of the large spatial correlation of the magnification \citep{CoorayHutererHolz2006}.

There is also strong evidence that high-redshift GRBs are significantly affected by gravitational lensing and magnification bias \citep[e.g.][]{PorcianiVielLilly2007,WyitheOhPindor2010_arXiv}, which may lead to large biases in their deduced distances if not properly accounted for. Moreover, magnification scatter will almost certainly be the dominant source of error in distance measurements using MSBs \citep{ArunEtal2009}.

While a source of noise for distance measurements, magnification of standard candles/sirens has also been considered a useful signal of the cosmic matter distribution.
The observed scatter in SNe brightnesses can be used to constrain the amount of dark matter in the form of massive compact objects \citep{Rauch1991,MetcalfSilk1999,MoertsellGoobarBergstroem2001,MintyHeavensHawkins2002}, or to  
measure the power spectrum of density fluctuations on small scales \citep{Metcalf1999} and on larger scales \citep{DodelsonVallinotto2006}. Correlations between galaxies and SN brigtnesses can be used to constrain the properties of galaxy halos \citep{Metcalf2001,JoenssonEtal2010_GOODS,JoenssonEtal2010_SNLS}.

When measuring distances to standard candles/sirens, one can try to correct for the gravitational lensing effects by inferring the magnification from other observations. For example, \citet{GunnarssonEtal2006} and \citet{JoenssonMoertsellSollerman2009} suggested modelling the matter distribution between the observer and a source using a model for the halos of the observed galaxies near the line of sight. These foreground models could reduce magnification scatter for high-redshift SN by up to a factor of 3. However, the success of this technique relies heavily on the validity of many assumptions about the detailed relation between observed light and total matter.

\citet{DalalEtal2003} used convergence power spectra to estimate the correlation between the true convergence (which is closely linked to the magnification) and the convergence as reconstructed from background galaxy shear measurements. They found that even with a very high number density of galaxies for the convergence reconstruction, the reduction in the magnification-induced scatter was small. Based on similar calculations, \citet{ShapiroEtal2010} estimated that convergence maps reconstructed using galaxy flexion in addition to shear might help to reduce the lensing-induced distance errors by up to 50\%, making this particular `delensing' approach promising.

In this work, we use ray-tracing through the Millennium Simulation \citep{SpringelEtal2005_Millennium} to investigate how accurately the lensing magnification of standard candles and standard sirens in a $\Lambda$CDM universe can be inferred from the shear and flexion measured in the images of distant galaxies. The accuracy with which the magnification can be estimated directly determines the lensing-induced uncertainty in the distances of type-Ia SNe, GRBs, and MSBs that remains after delensing, i.e. after a correction for the magnification effects has been applied. In a companion paper \citetext{Gair, King, \& Hilbert, in prep.}, we will discuss the consequences of delensing for the use of gravitational-wave sources as cosmological probes in comparison to other probes.

Compared to the magnification correction scheme suggested, e.g., by \citet{GunnarssonEtal2006}, the method discussed here has the advantage that it does not require any assumptions about the relation between the distribution of visible and dark matter. This method consists of (i) reconstructing the convergence toward a SN, GRB, or MSB as accurately as possible from the shear and flexion measurements of galaxy images near the line-of-sight, (ii) estimating the magnification from the reconstructed convergence, and (iii) correcting the observed SN, GRB, or MSB signal using the inferred magnification.

As will be shown below, for the correction to yield satisfactory results, one must reconstruct the convergence on sub-arc-minute scales \citep[see also][]{DalalEtal2003}, where the non-Gaussian nature of the underlying density field becomes important. Moreover, a realistic description of the noise of the convergence reconstruction on these scales has to take into account the discreteness and randomness of the galaxy image positions. Finally, the relation between the magnification and the noisy reconstructed convergence is non-trivial. Such complications usually have to be neglected in a treatment relying on convergence correlations and Gaussian statistics \citep[as was done, e.g, by ][]{DalalEtal2003, ShapiroEtal2010}. In contrast, all these effects can be included in the ray-tracing simulations used here. This approach not only provides more realistic predictions about the performance of magnification correction schemes, but also an optimised magnification estimate, and more detailed information about the residual magnification after correction. In particular, the simulations yield detailed probability distributions for the magnification and its residual as a function of the reconstructed convergence. Such information is valuable as input for Bayesian parameter estimation for cosmological models, and helps to further reduce lensing-induced errors and biases on cosmological parameters \citep[][]{HirataHolzCutler2010,ShangHaiman2010_arXiv}.

The paper is organised as follows: In Section~\ref{sec:theory}, we discuss gravitational lensing and how it affects the distance estimates of standard candles and standard sirens. In Section~\ref{sec:methods}, we describe the methods for simulating noise-free lensing maps and for creating shear and flexion maps with realistic noise properties. Moreover, we discuss how the information in the noisy maps is used to estimate the local magnification.
Results for the magnification correction are presented in Section~\ref{sec:results}. The paper concludes with a summary and discussion in Section~\ref{sec:summary}.

\section{Theory}
\label{sec:theory}

\subsection{Gravitational lensing}
\label{sec:lensing_basics}

The photons and gravitons (and any other particle species, e.g., neutrinos) from a distant source may be deflected gravitationally by intervening matter inhomogeneities before they reach the observer. The deflection, called gravitational lensing \citep[see, e.g.,][for an overview]{SchneiderKochanekWambsganss_book}, causes shifts in the observed image position relative to the `true' source position, i.e. the sky position at which the source would be seen in the absence of gravitational lensing. Here, the lensing effects are described by the lens mapping,
\begin{equation}
  \label{eq:lens_map}
 \mathcal{L}:\;\R^2\times[0,\infty) \to \R^2 :\; (\vect{\theta},z) \mapsto \vect{\beta} = \vect{\beta}(\vect{\theta}, z)
 ,
\end{equation}
which relates the angular image position $\vect{\theta}=(\theta_1,\theta_2)$ of a point-like source at redshift $z$ to its true (unlensed) angular source position $\vect{\beta}=(\beta_1,\beta_2)$.\footnote{
The flat-sky approximation is used here to avoid irrelevant complications due to the spherical geometry of the `real' sky, and is valid given the survey areas that we consider.}

The distortion induced in the images of extended sources like galaxies can be quantified to first order by the Jacobian of the lens mapping:
\begin{equation}
\label{eq:lens_distortion}
\begin{split}
\tens{A}(\vect{\theta}, z) &= \left(\parder{\beta_i(\vect{\theta}, z)}{\theta_j}\right)_{i,j=1,2}
\\&=
\begin{pmatrix}
1 - \kappa - \gamma_1 &   - \omega - \gamma_2 \\
    \omega - \gamma_2 & 1 - \kappa + \gamma_1 
\end{pmatrix}.
\end{split}
\end{equation}
This equation defines the convergence $\kappa(\vect{\theta},z)$, the shear $\gamma(\vect{\theta},z)= \gamma_1 + \ii \gamma_2$, and the asymmetry $\omega(\vect{\theta},z)$.
Higher-order image distortions can be quantified by higher derivatives of the lens mapping, which can be combined, e.g., into the spin-1 flexion
\begin{equation}
  \F(\vect{\theta},z) = \nablam \gamma(\vect{\theta},z),
\end{equation}
and the spin-3 flexion
\begin{equation}
  \G(\vect{\theta},z) = \nablap \gamma(\vect{\theta},z),
\end{equation}
where
\begin{equation}
  \nablapm=\left(\parder{}{\theta_1} \pm \ii \parder{}{\theta_2}\right).
\end{equation}
The (unsigned) image magnification is given by:
\begin{align}
  \mu(\vect{\theta}, z) &= \left|\det\tens{A}(\vect{\theta}, z)\right|^{-1}
\nonumber\\&=
\label{eq:mu_from_kappa_gamma_and_omega}
\left| (1-\kappa)^2 - |\gamma|^2 + \omega^2 \right|^{-1}
.
\end{align}

\subsection{Lensing effects on distance estimates}
\label{sec:distance_estimates}

The lensing magnification influences the observed signal strength of distant SN, GRBs, and MSBs, and thereby the inferred distances \citep[e.g.][]{Gunn1967b,SchneiderWagoner1987,KantowskiVaughanBranch1995, Frieman1996, WambsganssEtal1997}. In a nutshell, the radiation power $P_\mathrm{S}$ emitted by a standard candle/siren at redshift $\zS$ and the power $P_\mathrm{R}$ received by a detector are related by:\footnote{
In practice, received radiation energies or photon numbers $\propto P_\mathrm{R}$ are measured for type-Ia SN and GRBs, while 
gravitational-wave detectors record wave amplitudes $\propto \sqrt{P_\mathrm{R}}$. This difference does, however, not affect the resulting relation between inferred distances and magnification.
}
\begin{equation}
  P_\mathrm{R} =  \frac{c_\mathrm{R} \mu(\vect{\theta}, \zS)}{\Dlum^2(\zS)}P_\mathrm{S}.
\end{equation}
Here, $c_\mathrm{R}$ denotes a detector-specific constant, $\mu(\vect{\theta}, \zS)$ the lensing magnification in the direction $\vect{\theta}$ of the image, and $\Dlum$ the luminosity distance. Hence,
\begin{equation}
  \Dlum(\zS) = \sqrt{c_\mathrm{R} \mu(\vect{\theta},\zS) \frac{P_\mathrm{S}}{P_\mathrm{R}}}.
\end{equation}

An error $\delta \mu$ in the magnification causes an error $\delta \Dlum$ in the luminosity distance:
\begin{equation}
  \delta \Dlum \approx -\frac{1}{2} \frac{\delta \mu}{\mu} \Dlum,
\end{equation}
or in logarithmic form,
\begin{equation}
  \delta \ln(\Dlum) \approx -\frac{1}{2} \delta \ln(\mu).
\end{equation}

\subsection{Quantifying the magnification uncertainty}
\label{sec:quantifying_the_magnification_uncertainty}

Delensing methods provide estimates $\muest$ of the true magnification $\mu$ of a standard candle/siren. The resulting error of the delensing method is encoded in the distribution of the residual magnification $\mures = \mu/\muest$. A worthwhile delensing method should result in a residual magnification whose statistical dispersion is substantially smaller than the dispersion in the true magnification $\mu$.

In order to quantify the performance of delensing methods, one needs a way to quantify the dispersion in the residual magnification $\mures$ and compare it to the dispersion in the uncorrected magnification $\mu$. A common way to quantify the dispersion of a distribution is the standard deviation. However, for point-like sources, the magnification distribution $p_{\mu}(\mu)\propto\mu^{-3}$ for $\mu\gg 1$ and so the standard deviation of the magnification
is formally divergent. In this work we will therefore quantify the dispersion of the magnification using the distribution of the logarithm of the magnification
\begin{equation}
\lambda=\ln(\mu)
\end{equation}
 and its standard deviation, $\sigma_{\lambda}$, which is finite for point-like sources. Similarly,  the dispersion of the residual magnification $\mures$ will be quantified by the standard deviation $\sigma_{\lambdares}$ of its logarithm
\begin{equation}
\lambdares=\ln(\mures).
\end{equation}

\subsection{Magnification estimates from galaxy shear and flexion}
\label{sec:reconstruction_basics}

There are various methods to obtain an estimate for the magnification $\mu(\vect{\theta},z)$ for a given line of sight $\vect{\theta}$ and redshift $z$. One can, for example, try to infer the matter distribution near that line of sight from the observed distribution of mass tracers like galaxies, and then calculate the expected magnification $\mu(\vect{\theta},z)$ from the inferred matter distribution \citep[e.g.][]{GunnarssonEtal2006}. One can also try to infer the magnification from source number density counts \citep[e.g.][]{BroadhurstTaylorPeacock1995, ZhangPen2005_CM}, or the image sizes of extended sources \citep[e.g.][]{BartelmannNarayan1995}.

Here, we focus on methods that use weak-lensing reconstruction \citep{KaiserSquires1993,BaconEtal2006} to obtain magnification estimates \citep[see also][]{DalalEtal2003,ShapiroEtal2010}. These methods `reconstruct' the convergence from the shear and/or flexion measured in the images of distant galaxies and use the result to calculate the magnification. In their conventional form, they neglect the asymmetry, and assume that the convergence, shear, and flexion are well approximated by derivatives of a common lens potential $\psi(\vect{\theta},z)$:\footnote{
These approximations are exact for pure E-mode lensing. B-mode lensing can be taken into account by taking a second potential into consideration \citep[e.g.][]{SchneiderVanWaerbekeMellier2002,HirataSeljak2003}. B-modes are, however, usually very small compared to E-modes \citep[][]{JainSeljakWhite2000,HilbertEtal2009_RT}.
}
\begin{subequations}
\begin{align}
 \omega(\vect{\theta},z) &= 0,
\\
 \kappa(\vect{\theta},z) &= \frac{1}{2} \nablap \nablam               \psi(\vect{\theta},z),
\\
 \gamma(\vect{\theta},z) &= \frac{1}{2} \bigl(\nablap\bigr)^2         \psi(\vect{\theta},z),
\\
     \F(\vect{\theta},z) &= \frac{1}{2} \bigl(\nablap\bigr)^2 \nablam \psi(\vect{\theta},z),
\\
     \G(\vect{\theta},z) &= \frac{1}{2} \bigl(\nablap\bigr)^3         \psi(\vect{\theta},z)
.
\end{align}
\end{subequations}
Hence, the convergence can be reconstructed from the shear and flexion, e.g., by using the relations
\begin{subequations}
\begin{align}
\label{eq:ft_kappa_from_gamma}
  \hat{\kappa}(\vect{\ell},z) &= \frac{(\ell^*)^2}{|\ell|^2} \hat{\gamma}(\vect{\ell},z),
\\
 \label{eq:ft_kappa_from_F}
  \hat{\kappa}(\vect{\ell},z) &= -\ii \frac{\ell^*}{|\ell|^2} \hat{\F}(\vect{\ell},z) \text{, and}
\\
\label{eq:ft_kappa_from_G}
  \hat{\kappa}(\vect{\ell},z) &= -\ii \frac{(\ell^*)^3}{|\ell|^4} \hat{\G}(\vect{\ell},z)
\end{align}
\end{subequations}
in Fourier space. Here, hats denote Fourier transforms, i.e.
\begin{equation*}
  \hat{\kappa}(\vect{\ell},z) = \frac{1}{2\pi}\int_{\R^2}\!\diff[2]{\vect{\theta}}\,\exp(-\ii \vect{\ell}\cdot\vect{\theta}) \kappa(\vect{\theta},z),
\end{equation*}
$\vect{\ell}=(\ell_1,\ell_1)$ denotes the two-dimensional Fourier vector, $\ell = \ell_1 + \ii \ell_2$ denotes the complex Fourier wave number, and an asterisk denotes complex conjugation.\footnote{The definition of the Fourier transform used here differs from the one used, e.g., by \citet{KaiserSquires1993}.}
The reconstructed convergence is then translated into a magnification estimate via an approximate version of Eq.~\eqref{eq:mu_from_kappa_gamma_and_omega} applicable in the limit of weak lensing:
\begin{subequations}
\begin{align}
\label{eq:mu_from_kappa}
  \muest(\vect{\theta}, z) &\approx \bigl[1-\kappa(\vect{\theta}, z)\bigr]^{-2}
\\
&\approx 1 + 2 \kappa(\vect{\theta}, z)
.
\end{align}
\end{subequations}
The magnification estimate $\muest(\vect{\theta}, z)$ is then used to `delens' the standard candle/siren.

There are several complications in practice. First, neither the shear $\gamma$, nor the flexion $\F$ and $\G$ can be directly measured, but only their reduced versions $g=\gamma/(1-\kappa)$, $F=\nablam g$, and $G = \nablap g$ \citep[][]{SchneiderSeitz1995_I, SchneiderEr2008}. This complication can be taken into account, e.g., by iterative reconstruction methods \citep[][]{SchneiderSeitz1995_II}. Moreover, $\gamma\approx g$, $\F\approx F$, and $\G \approx G$ for most parts of the sky. It is thus assumed in the following that galaxy shapes provide estimates for the shear and flexion (and we postpone the proper treatment of reduced shear and flexion to forthcoming studies).

Even from a perfect shear or flexion map, the convergence can only be reconstructed up to a constant because of the mass-sheet degeneracy \citep[][]{FalcoGorensteinShapiro1985}. In addition, the boundary conditions influence the reconstruction near the edges of finite fields. These problems can, however, be alleviated by using a large field of view for the reconstruction \citep[whose mean convergence is close enough to zero, as discussed by][]{ShapiroEtal2010} and special finite-field inversion techniques \citep[e.g.][]{SeitzSchneider1996}.

Furthermore, the galaxies that can be used to measure the shear and flexion have a finite number density $\ngal$ and thus only provide shear and flexion estimates at a finite number of discrete and randomly distributed points. Moreover, the shape of a galaxy image provides a very noisy estimate for the shear and flexion. Therefore, spatial interpolation and averaging are necessary to obtain a convergence estimate with acceptable signal-to-noise in the desired direction of a standard candle/siren. The smoothing erases real structure on small scales, which causes additional deviations of the reconstructed convergence from the `true' convergence. The optimal smoothing scale is determined by a trade-off between the aim to reconstruct as much small-scale structure of the convergence as possible and to average over as many galaxy images as possible, so as to reduce the noise due to the dispersion in their intrinsic shapes \citep[][]{DalalEtal2003}.

Finally, the galaxies do not all lie at the redshift $\zS$ of the standard candle/siren, but have a certain redshift distribution $\pgal(\zgal)$ depending on their intrinsic redshift distribution and on the depth of the survey. 
A reconstruction scheme not using individual galaxy redshifts\footnote{
Taking into account the individual galaxy redshifts (e.g. in a tomographic reconstruction) complicates the reconstruction, but may improve the magnification estimates. A simple way of using the individual redshifts to improve the reconstruction is discussed in Section~\ref{sec:redshift_weighting}.}
 only yields a noisy and smoothed version of the effective convergence:
\begin{equation}
	\kappaeff(\vect{\theta}) = \int\diff{\zgal}\, \pgal(\zgal)\, \kappa(\vect{\theta},\zgal).
\end{equation}
If the redshift distribution $\pgal(\zgal)$ is not sharply peaked around the redshift $\zS$ of the standard candle/siren, the effective convergence $\kappaeff(\vect{\theta})$ may deviate substantially from the convergence $\kappa(\vect{\theta},\zS)$, which may cause a systematic as well as a statistical error in the reconstruction of $\kappa(\vect{\theta},\zS)$. Spatial variations of the `local' redshift distribution of the galaxies constitute an additional source of noise.

\section{Methods}
\label{sec:methods}

The complications discussed above limit the accuracy with which the magnification can be estimated from the weak-lensing reconstruction. Therefore, accurate estimates of their effects are needed in order to evaluate the performance of the weak-lensing reconstruction. Since it is difficult to obtain accurate estimates for all these effects by analytical calculations, we use a numerical approach.

Using ray-tracing through structure-formation simulations, we create a suite of simulated fields of view populated with galaxy images at random positions with noisy estimates of the local shear and flexion. The simulated galaxy images are then used to reconstruct the convergence in the simulated fields, and the reconstructed convergence is used to calculate an estimate of the magnification. The `true' magnification (obtained from the ray-tracing) in the field is then corrected by the magnification estimate from the reconstruction, and the resulting residual magnification is analysed.

\subsection{The ray-tracing}
\label{sec:ray_tracing}

Here, the Millennium Simulation (MS) by \citet{SpringelEtal2005_Millennium}, a large high-resolution $N$-body simulation of cosmic structure formation, is used to simulate the matter distribution between observer and source.
The cosmological parameters for the MS are: a matter density $\Omega_\mathrm{M}=0.25$ and a cosmological-constant energy density $\Omega_\Lambda=0.75$ (in units of the critical density), a Hubble constant $h=0.73$ (in units of $100\,\mathrm{km}\,\mathrm{s}^{-1}\Mpc^{-1}$), a primordial spectral index $n=1$ and a normalisation parameter $\sigma_8=0.9$ for the linear density power spectrum.
The simulation was run using a parallel TreePM version of \textsc{GADGET2} \citep{Springel2005_GADGET2} with $N_\mathrm{p}=2160^3$ particles of mass $m_\mathrm{p}=8.6\times10^8h^{-1}\,\Msolar$ in a cubic box with comoving side length $L=500h^{-1}\,\Mpc$, and an effective force softening length $\epsilon=5h^{-1}\,\kpc$.

The gravitational lensing effects from the matter structures in the MS are calculated by the Multiple-Lens-Plane ray-tracing algorithm described in \citet[][]{HilbertEtal2009_RT}.
The algorithm takes into account the lensing effects of the dark-matter structures on scales $\gtrsim5h^{-1}\,\kpc$, which are represented by the simulation particles in the MS. In addition, the effects of the stars in galaxies with stellar masses $\geq10^9h^{-1}\,\Msolar$ (as predicted by the model of \citet[][]{DeLuciaBlaizot2007}) are taken into account as described in \citet[][]{HilbertEtal2008_StrongLensing_II}.
 
The multi-plane algorithm is used to simulate 32 fields of view, each with an area of $4\,\degt\times4\,\degt$ (providing $512\,\degt^2$ total area) and covered with a regular grid of $4096^2$ pixels (providing a resolution of $3.5\,\arcsect$) in the image plane. For each pixel, a ray is traced back through the MS by the algorithm to calculate the magnification, convergence, shear, and flexion for sources on each of the 44 lens planes that span the redshift range from $z=0$ to $z=5.7$.

Every pixel of the ray-tracing fields (excluding a $15\,\arcmint$ margin, which might be affected by artefacts in the weak-lensing reconstruction, and strong-lensing regions with $\det(\tens{A}) \leq 0$ or $\tr(\tens{A}) \leq 0$) is used as a sample line-of-sight towards a standard candles/siren. Here, we consider only volume-limited samples of standard candles/sirens. Thus, each pixel is weighted in the statistical analysis by its inverse magnification to account for the magnification bias \citep[][]{Canizares1981, HilbertEtal2007_StrongLensing}.

\subsection{The simulated weak-lensing surveys}
\label{sec:mock_catalogs}

Simulated catalogues of lensed galaxy images with a given number density $\ngal$ are generated with random sky positions uniformly distributed in the fields of view. The redshifts of the simulated galaxies are drawn from a distribution that matches observations \citep[][]{BrainerdBlandfordSmail1996}
\begin{equation}
\label{chap:p_gal}
\pgal(\zgal)= \frac{\beta}{z_0\,\Gamma\!\left(\frac{\alpha+1}{\beta}\right)} \left(\frac{\zgal}{z_0}\right)^{\alpha}\exp\left[-\left(\frac{\zgal}{z_0}\right)^{\beta} \right],
\end{equation}
where $\alpha$, $\beta$, and $z_0$ are parameters (with $z_0$ proportional to the median redshift of the simulated survey). 

Here, we consider three idealised weak-lensing galaxy surveys: an \emph{advanced survey}, a \emph{futuristic survey}, and a \emph{perfect survey}. For the advanced survey, we assume a galaxy density $\ngal = 100\,\arcmint^{-2}$ and a redshift distribution  with $\alpha = 2$, $\beta = 3/2$, and median redshift $\zmedian = 1.5$ appropriate for a deep space-based lensing survey \citep[cf., e.g.,][]{SchrabbackEtal2010}. For the futuristic survey, we assume a very high density $\ngal = 500\,\arcmint^{-2}$ of usable galaxy images having a redshift distribution with $\alpha = 0.8$, $\beta = 2$, and $\zmedian = 1.8$ appropriate for an ultra-deep survey \citep[][]{CoeEtal2006} with extremely high spatial resolution.\footnote{
Even with future telescopes, such high densities of galaxies with shape measurements might only be reached for a very deep pointed observation, which could then be supplemented by a wider survey with lower number densities \citep[][]{ShapiroEtal2010}. 
}
In addition, we consider a `perfect' survey, whose noise-free convergence maps are directly created from the ray-tracing.

For the advanced and futuristic surveys, we consider both the case that only shear estimates are used, and the case that shear estimates are combined with flexion estimates. The shear and flexion estimates of the simulated galaxies are obtained from interpolating the shear and flexion from the ray-tracing onto the galaxy positions and redshifts. The shape noise in each galaxy image is simulated by adding a random number drawn from a normal distribution with standard deviation $\sigmagammagal = 0.2$, $\sigmaFgal = 0.5\,\arcmint^{-1}$, and $\sigmaGgal = 0.9\,\arcmint^{-1}$, to each component of $\gamma$, $\F$, and $\G$ respectively \citep[Rowe et al., unpublished, see][]{ShapiroEtal2010}.

\subsection{The weak-lensing reconstruction}
\label{sec:reconstruction}

For the weak-lensing reconstruction, we consider a mesh-based method that allows one to employ Fast Fourier Transforms (FFT). Each field of view is covered with a regular mesh of $4096^2$ pixels (i.e. the same mesh geometry as for the ray-tracing). In each pixel, the shear $\gamma$ is estimated by weighted sums over the shear estimates $\gammagali$ of the simulated galaxy images in the field:
\begin{equation}
\label{eq:projection_sum}
  \gamma = \frac{ \sum_{i} \wgali \gammagali }{\sum_{i} \wgali}
  .
\end{equation}
The weights $\wgali$ are determined by:\footnote{See Section~\ref{sec:redshift_weighting} for using redshift-dependent weights.}
\begin{equation}
  \wgali = w_\theta(\vect{\theta} - \thetagali)
\end{equation}
where $\vect{\theta}$ denotes the sky position of the pixel centre, the $\thetagali$ denote the sky positions of the galaxy images,
\begin{equation}
  w_\theta(\theta) = \exp\left(-\frac{|\vect{\theta}|^2}{2 \thetaw^2}\right),
\end{equation}
and $\thetaw = 1.75\,\arcsect$ denotes a weighting scale, which is chosen slightly smaller than the pixel size. The shear estimate on the regular mesh is then Fourier transformed by FFT methods \cite[][]{FrigoJohnson2005_FFTW3}, and converted into a convergence estimate $\hat{\kappa}_{\gamma}$ using Eq.~\eqref{eq:ft_kappa_from_gamma} and setting $\hat{\kappa}_{\gamma}(\vect{\ell}=\vect{0}) = 0$. If only shear information is used, $\hat{\kappa}_{\gamma}$ is used directly as an estimate of the `raw' convergence in Fourier space prior to additional smoothing:
\begin{equation}
  \hatkapparaw(\vect{\ell}) = \hat{\kappa}_{\gamma}(\vect{\ell}).
\end{equation}

In the case that flexion information is used as well, estimates $\hat{\kappa}_{\F}$ and $\hat{\kappa}_{\G}$ of the Fourier-space convergence from the galaxy flexion $\F$ and $\G$ are computed in a procedure analogous to that for $\hat{\kappa}_{\gamma}$. The convergence estimates from shear and flexion are then combined into a joint estimate by \citep[][]{OkuraUmetsuFutamase2007}:
\begin{equation}
 \hatkapparaw(\vect{\ell}) = 
\frac{\sigmagammagal^{-2} \hat{\kappa}_{\gamma}(\vect{\ell}) + |\ell|^2\sigmaFgal^{-2} \hat{\kappa}_{\F}(\vect{\ell}) + |\ell|^2\sigmaGgal^{-2} \hat{\kappa}_{\G}(\vect{\ell})}{\sigmagammagal^{-2} + |\ell|^2\sigmaFgal^{-2} + |\ell|^2\sigmaGgal^{-2}}
.
\end{equation}

The smoothing inherent in the averaging \eqref{eq:projection_sum} may be insufficient to reach an acceptable signal-to-noise ratio of the convergence estimate in real space. The raw convergence estimate
$\hatkapparaw$ is thus smoothed by a Gaussian low-pass filter with filter scale $\thetas$, i.e.
\begin{equation}
\label{eq:Gauss_filter}
 \hatkappaest(\vect{\ell}) = \exp\left(-\frac{\thetas^2}{2}|\vect{\ell}|^2\right) \hatkapparaw(\vect{\ell}),
\end{equation}
before the resulting convergence estimate is transformed back to real space by FFT methods.\footnote{See Section \ref{sec:Wiener_filters} for using Wiener filters instead of Gaussian smoothing.}

\subsection{Magnification estimates from noisy convergence maps}
\label{sec:magnification_from_convergence}

The convergence map generated from the weak-lensing reconstruction can be converted into a map of estimated magnification in several ways. If the estimated convergence $\kappaest(\vect{\theta})$ is close to true convergence $\kappa(\vect{\theta},\zS)$, and if the shear and rotation are small,
\begin{equation}
\label{eq:simple_magnification_estimate}
	\lambdaest(\vect{\theta}) = -2\ln\bigl[1 - \kappaest(\vect{\theta})\bigr]
\end{equation}
provides a good estimate for the logarithmic magnification $\lambda(\vect{\theta}, \zS)$ of a standard candle/siren at position $\vect{\theta}$ and redshift $\zS$.\footnote{
One might consider using the full Eq.~\eqref{eq:mu_from_kappa_gamma_and_omega} instead, but tests show that this does not improve the magnification estimate for realistic noise levels in the reconstruction (except near strong-lensing regions where the discussed reconstruction method is not suitable anyway).
}
This simple estimate might however fail, if the estimated convergence $\kappaest(\vect{\theta})$ differs substantially from the convergence $\kappa(\vect{\theta},\zS)$, e.g. because of shape noise or because the weak-lensing survey probes a redshift range very different from the redshift $\zS$ of the standard candle/siren. In certain cases, the simple estimate may perform even worse than simply ignoring lensing and assuming $\lambdaest = 0$ (see Section~\ref{sec:realistic_reconstruction}).

If the conditional distribution $p_{\lambda|\kappaest}(\lambda|\kappaest)$ of the true magnification $\lambda$ for a given estimated convergence $\kappaest$ is known, one can construct a bias-free magnification estimate that minimises the dispersion $\sigma_{\lambdares}$ in the residual magnification by (see Appendix~\ref{sec:app_magnification_from_convergence}):
\begin{equation}
\label{eq:optimal_magnification_estimate}
  \lambdaest(\vect{\theta}) =  \EV{\lambda}_{\lambda|\kappaest}\bigl[\kappaest(\vect{\theta})\bigr],
\end{equation}
where $\EV{\lambda}_{\lambda|\kappaest}(\kappaest)$ is the expectation value of the true magnification $\lambda$ for a given estimated convergence $\kappaest$:
\begin{equation}
 \EV{\lambda}_{\lambda|\kappaest}(\kappaest) = \int \diff[]{\lambda}\, p_{\lambda|\kappaest}(\lambda | \kappaest) \lambda
.
\end{equation}
This estimate can be computed from the lensing simulations by spatial averages over the simulated fields.
The estimate is optimal in the sense that no other magnification estimate based on the estimated convergence yields a smaller dispersion in the residual magnification. For example, the resulting dispersion in the residual magnification is never larger than the dispersion in the uncorrected magnification (i.e. assuming $\lambdaest=0$).

\section{Results}
\label{sec:results}

\subsection{Perfect reconstruction}
\label{sec:perfect_reconstruction}

We first consider the case that one can somehow obtain a perfect estimate of the true convergence $\kappa(\vect{\theta},\zS)$ for sources at the same position $\vect{\theta}$ and redshift $\zS$ as the standard candle/siren.
This case provides hard limits to what can be gained by delensing in more realistic cases.

\begin{figure}
\centerline{\includegraphics[width=1\linewidth]{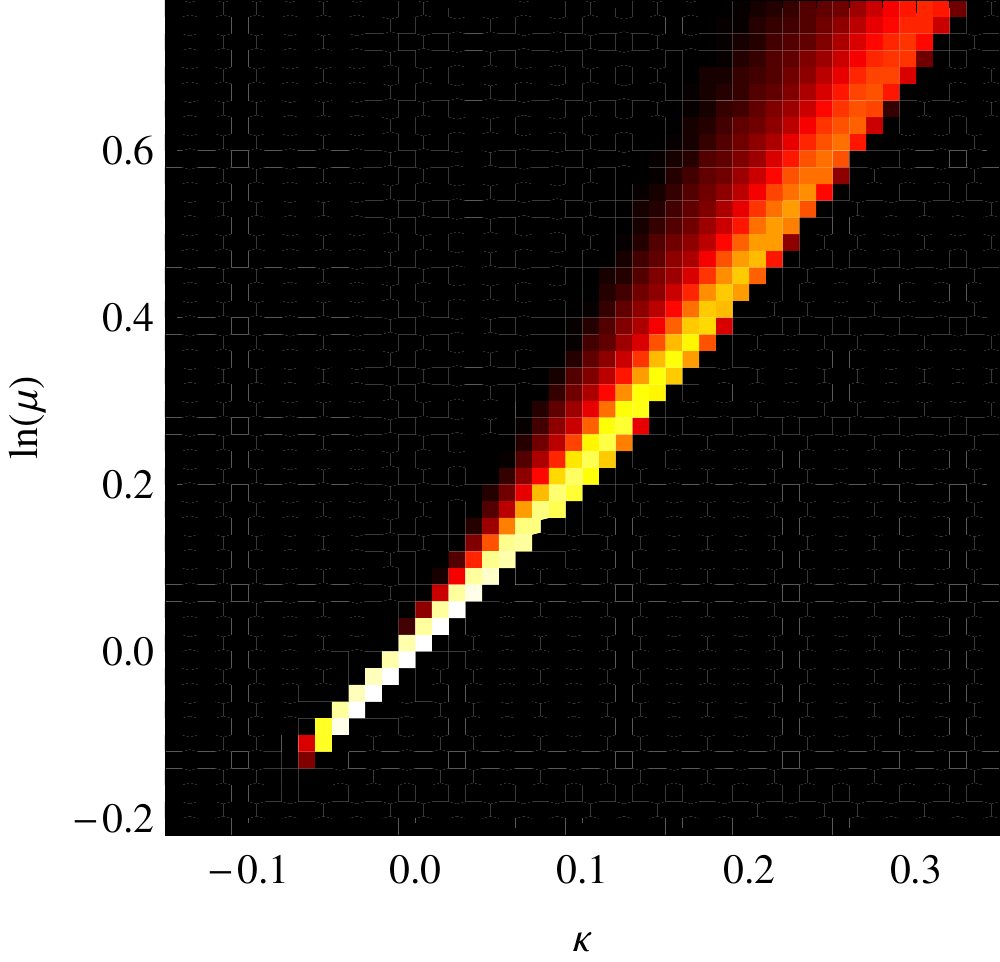}}
\caption{
\label{fig:perfect_reconstruction_p_mu_kappa}
The joint distribution of the logarithmic magnification $\lambda=\ln(\mu)$ and the convergence $\kappa$ for sources at redshift $\zS = 1.5$ (lighter areas correspond to higher probability densities on a logarithmic scale).
}
\end{figure}

The joint distribution $p_{\lambda,\kappa}(\mu,\kappa)$ of the logarithmic magnification $\lambda=\ln(\mu)$ and convergence $\kappa$ for sources at redshift $\zS = 1.5$ is shown in Fig.~\ref{fig:perfect_reconstruction_p_mu_kappa}. The joint distribution is almost degenerate with the probability concentrated around the relation $\lambda=-2\ln(1-\kappa)$. This indicates that one can get a very good magnification estimate from the  Eq.~\eqref{eq:optimal_magnification_estimate}, if the convergence can be reconstructed with sufficient accuracy. Moreover, the optimal estimate~\eqref{eq:optimal_magnification_estimate} is virtually identical to the simple estimate~\eqref{eq:simple_magnification_estimate}.

\begin{figure}
\centerline{\includegraphics[width=1\linewidth]{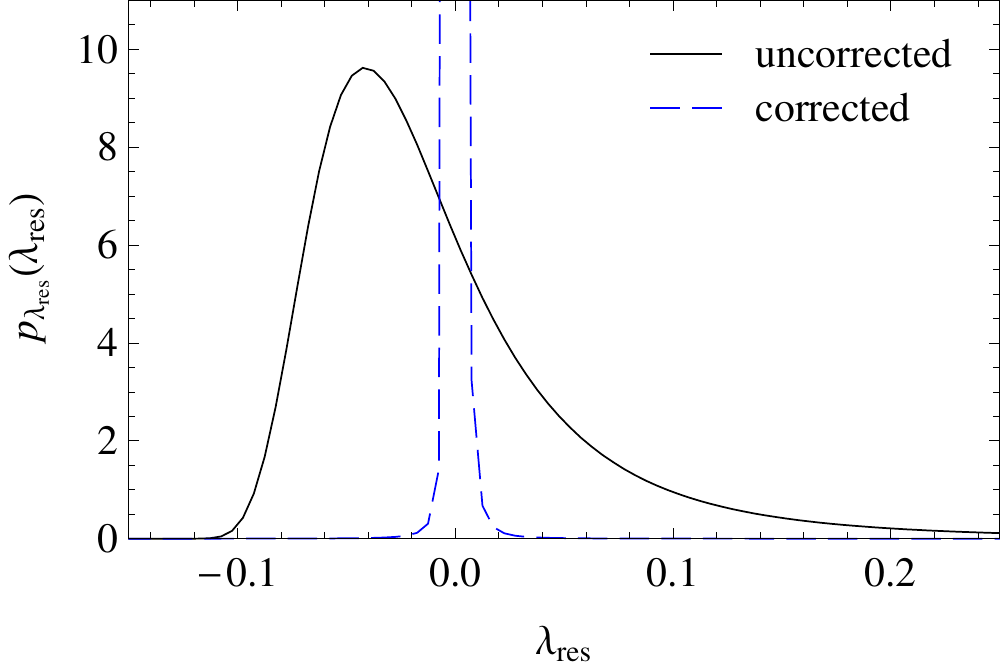}}
\caption{
\label{fig:perfect_reconstruction_p_mu}
The probability distribution $p_{\lambdares}(\lambdares)$ of the residual logarithmic magnification $\lambdares = \lambda - \lambdaest$ for sources at redshift $\zS = 1.5$.
The residual distribution for the optimal estimate~\eqref{eq:optimal_magnification_estimate} with the convergence to the same source redshift (dashed line) is compared to the case without correction (i.e. $\lambdaest=0$, solid line).
}
\end{figure}

The effect of delensing using a perfect convergence estimate is illustrated in Fig.~\ref{fig:perfect_reconstruction_p_mu}. The distribution of the uncorrected residual $\lambdares = \lambda$ for standard candles/sirens at redshift $\zS=1.5$ has a standard deviation $\sigma_{\lambdares}=0.075$. The corrected residual $\lambdares = \lambda - \lambdaest$ using the optimal estimate~\eqref{eq:optimal_magnification_estimate} and the convergence $\kappaest(\vect{\theta})=\kappa(\vect{\theta},\zS)$ has a fifteen times smaller standard deviation $\sigma_{\lambdares}=0.005$. The small but non-vanishing residual dispersion is due to the fact that the magnification also depends on the shear, but this dependence is ignored in the estimate \eqref{eq:optimal_magnification_estimate}.\footnote{
Tests show that taking the shear dependence into account does not improve the magnification estimate noticeably in the presence of realistic noise. We thus refrain from including an explicit shear dependence.
}

\begin{figure}
\centerline{\includegraphics[width=1\linewidth]{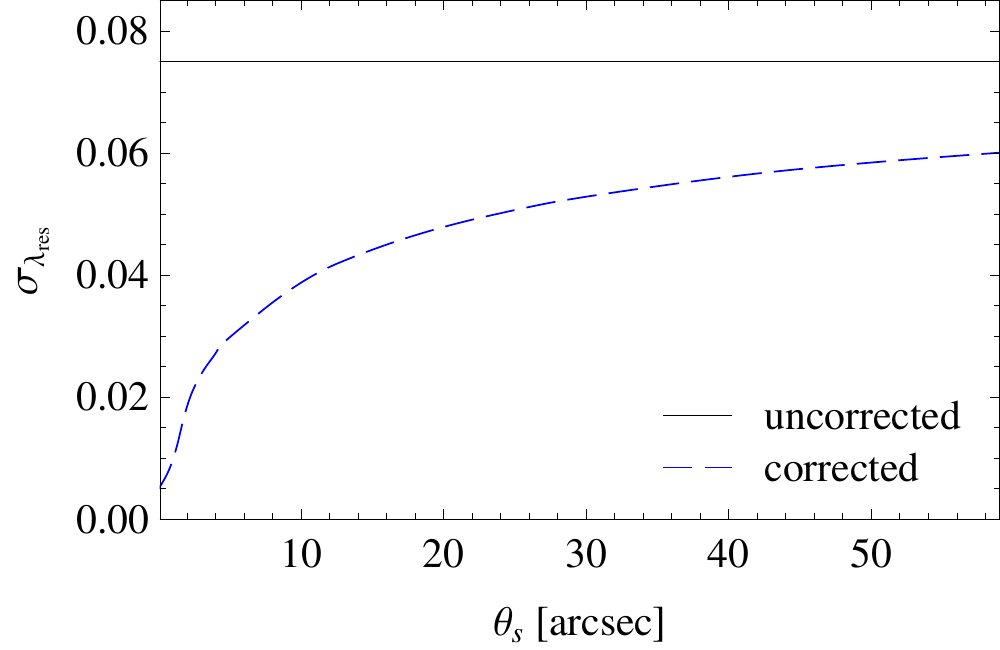}}
\caption{
\label{fig:perfect_reconstruction_residual_dispersion_and_smoothing}
The dispersion $\sigma_{\lambdares}$ of the residual $\lambdares$ (dashed line) for sources at redshift $\zS=1.5$ as a function of the filter scale $\thetas$ used to generate the convergence estimate $\kappaest$ by convolving the convergence $\kappa(\zS)$ with a Gaussian filter. The solid horizontal line marks the magnification dispersion $\lambdares=\lambda$ without correction.
}
\end{figure}

Even with perfect measurements for individual galaxies, a magnification estimate will make use of galaxies spread over a certain patch of the sky. This amounts to averaging or smoothing the convergence over the area occupied by the galaxies. The impact of smoothing on the dispersion of the residual magnification is shown Fig.~\ref{fig:perfect_reconstruction_residual_dispersion_and_smoothing}. The magnification correction is already substantially degraded (reaching 50\% of the uncorrected dispersion), if a smoothed version of the convergence with a filter scale $\thetas = 10\,\arcsect$ is used. This illustrates the need to faithfully reconstruct the convergence on scales of a few arc seconds, if one wants to decrease the dispersion in the residual magnification by a factor two compared to the uncorrected magnification.

\subsection{Realistic reconstruction}
\label{sec:realistic_reconstruction}

\begin{figure}
\centerline{\includegraphics[width=1\linewidth]{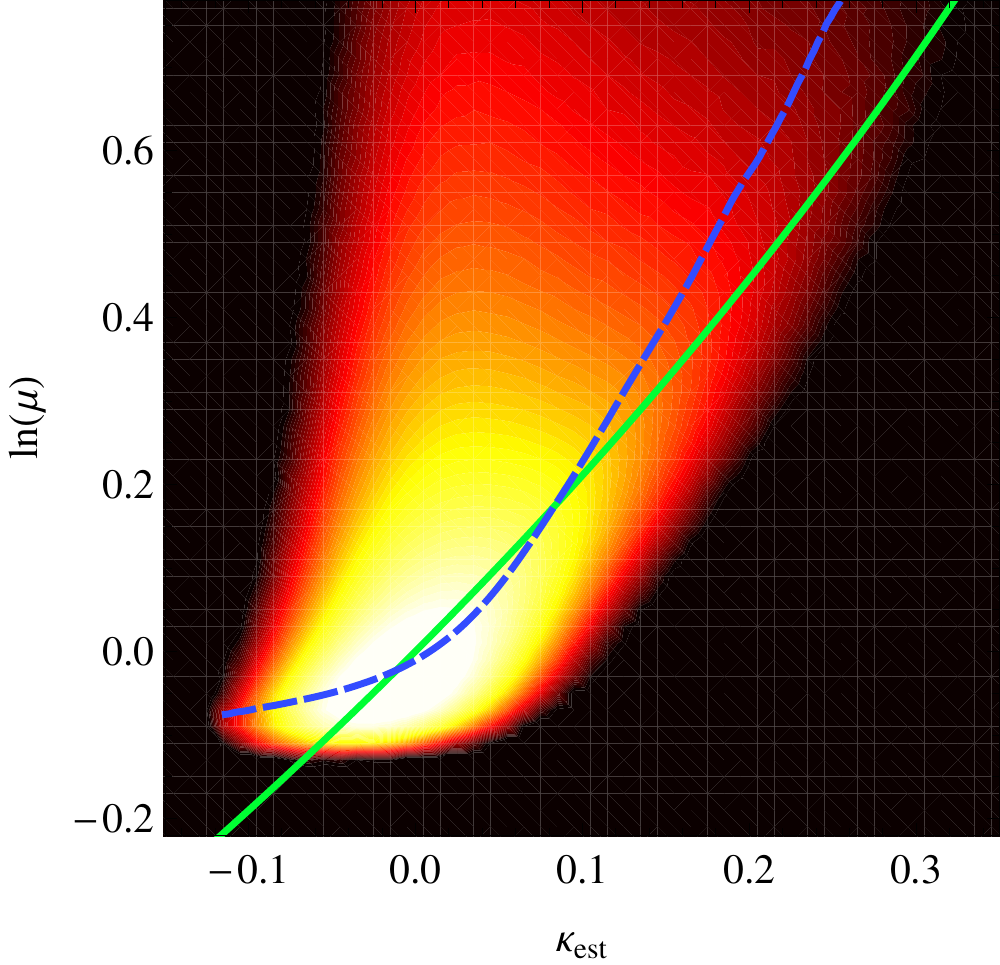}}
\caption{
\label{fig:advanced_reconstruction_p_mu_kappaest}
The joint distribution $p_{\lambda,\kappaest}(\lambda,\kappaest)$ of the logarithmic magnification $\lambda$ for sources at redshift $\zS = 1.5$ and the convergence $\kappaest$ reconstructed from the shear of galaxy images in a survey with galaxy density $\ngal=100\,\arcmint^{-2}$, median redshift $\zmedian = 1.5$, and a Gaussian smoothing with filter scale $\theta_\mathrm{s}=20\,\arcsect$ (lighter areas correspond to higher probability densities on a logarithmic scale).
Also shown are the relation $\lambda=-2\ln(1-\kappaest)$ (solid line) and the relation $\lambda = \EV{\lambda}_{\lambda|\kappaest}(\kappaest)$ (dashed line).
}
\end{figure}

As a more realistic scenario, we assume that the convergence $\kappaest$ is reconstructed from the shear measured in an advanced survey with a galaxy density $\ngal=100\,\arcmint^{-2}$ and median redshift $\zmedian = 1.5$. The joint distribution of the logarithmic magnification $\lambda$ for sources at redshift $\zS=1.5$ and the estimated convergence $\kappaest$ is shown in Fig.~\ref{fig:advanced_reconstruction_p_mu_kappaest}. The joint distribution roughly follows the relation	$\lambdaest = -2\ln(1 - \kappaest)$, but there is a large scatter around it. Moreover, the mean $\EV{\lambda}_{\lambda|\kappaest}(\kappaest)$ as a function of the estimated convergence $\kappaest$ deviates substantially from that relation, in particular for $\kappa \lesssim 0$. This shows that the estimate~\eqref{eq:simple_magnification_estimate} may differ substantially from the optimal estimate~\eqref{eq:optimal_magnification_estimate} for a realistic noise level.

\begin{figure}
\centerline{\includegraphics[width=1\linewidth]{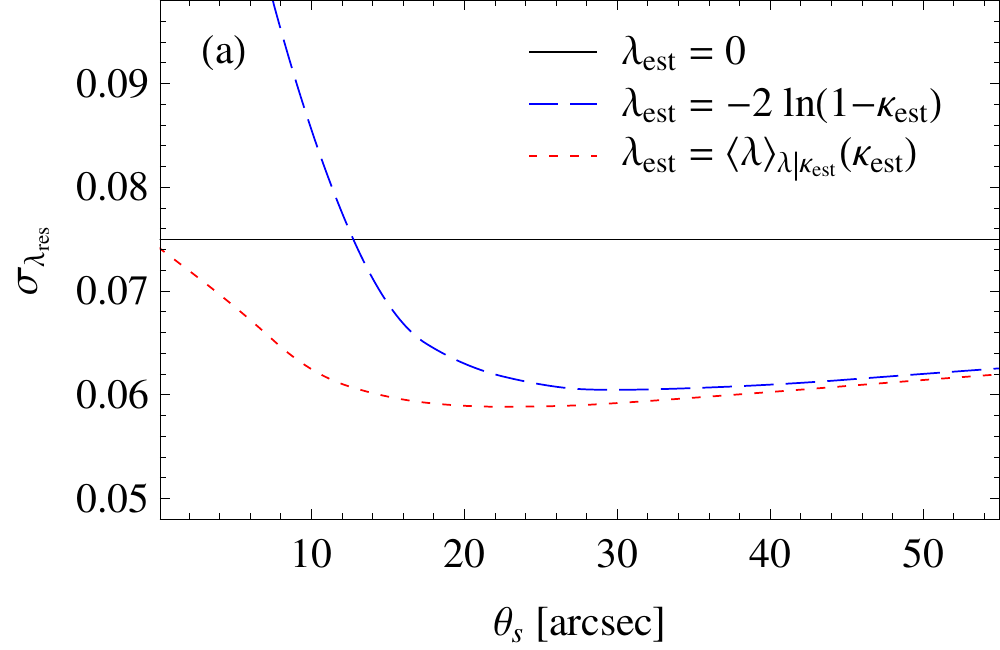}}
\centerline{\includegraphics[width=1\linewidth]{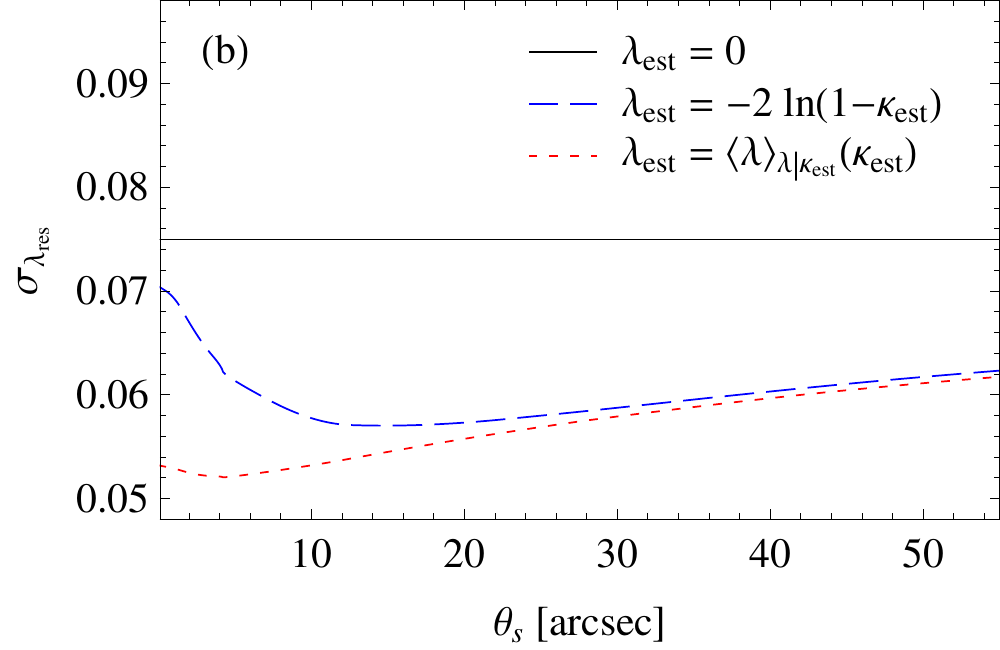}}
\caption{
\label{fig:advanced_reconstruction_residual_dispersion_and_smoothing}
The dispersion $\sigma_{\lambdares}$ of the residual logarithmic magnification $\lambdares = \lambda - \lambdaest$ of sources at redshift $\zS=1.5$ as a function of the filter scale $\thetas$ used to reconstruct the convergence $\kappaest$ from an advanced survey with $\ngal=100\,\arcmint^{-2}$. Results are shown for the simple estimate~\eqref{eq:simple_magnification_estimate} (dashed lines) and for the optimal estimate~\eqref{eq:optimal_magnification_estimate} (dotted lines) with convergence reconstructed from shear (a) and from shear and flexion (b). The solid horizontal line indicates the dispersion $\lambdares=\lambda$ without correction.
}
\end{figure}

Fig.~\ref{fig:advanced_reconstruction_residual_dispersion_and_smoothing} illustrates how much the dispersion $\sigma_{\lambdares}$ in the residual magnification $\lambdares$ of a standard candle/siren at $\zS=1.5$ can be reduced with an advanced survey for particular choices of the filter scale $\thetas$. For $\thetas \geq 30\,\arcsect$, both estimates \eqref{eq:simple_magnification_estimate} and \eqref{eq:optimal_magnification_estimate} yield a similar reduction in the residual dispersion. For $\thetas < 30\,\arcsect$, however, the optimal estimate \eqref{eq:optimal_magnification_estimate} performs significantly better than the simple estimate \eqref{eq:simple_magnification_estimate}.

The optimal estimate always yields a residual dispersion $\sigma_{\lambdares}$ smaller than the uncorrected dispersion $\sigma_{\lambda}$, but the simple estimate results in $\sigma_{\lambdares}>\sigma_{\lambda}$ for $\thetas \lesssim 10\,\arcsect$ if only shear data is used. In that case, the noise in the reconstructed convergence is much larger than the signal (i.e. the true convergence). The simple estimate translates this noise into a large noise in the estimated magnification, which thus acquires a dispersion exceeding the dispersion of the true magnification. In contrast, the optimal magnification estimate responds more weakly to the noisy convergence and thus keeps the residual dispersion below the uncorrected dispersion.

If only shear data is available, the residual dispersion $\sigma_{\lambdares}$ for sources at $\zS=1.5$ is reduced to $81\%$ of the uncorrected dispersion $\sigma_{\lambda}$, when the simple estimate and an optimal filter scale $\thetas \approx 30\,\arcsect$ is used. The optimal estimate yields a slightly smaller residual dispersion $\sigma_{\lambdares} = 0.79 \sigma_{\lambda}$ at the optimal filter scale $\thetas \approx 25\,\arcsect$. Further improvement is obtained if flexion information is available in addition to the shear. In this case, the residual dispersion is reduced to $76\%$ by the simple estimate and a filter scale $\thetas \approx 15\,\arcsect$. The optimal estimate \eqref{eq:optimal_magnification_estimate} performs best at $\thetas \approx 5\,\arcsect$, i.e. at the mean separation of the galaxy images. The residual dispersion is then $70\%$ of the uncorrected dispersion.

\begin{figure}
\centerline{\includegraphics[width=1\linewidth]{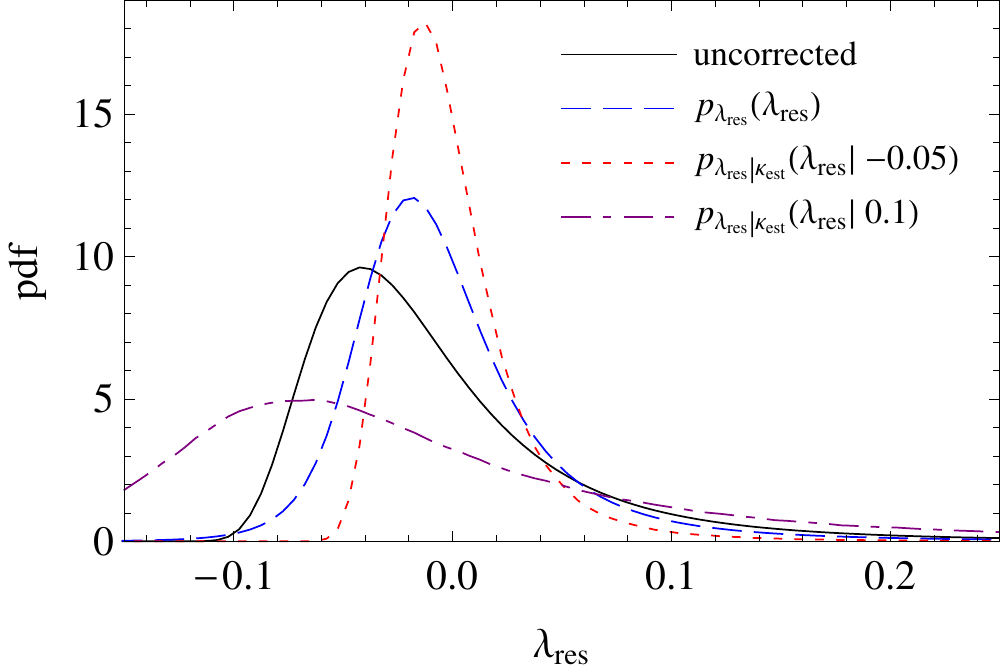}}
\caption{
\label{fig:advanced_reconstruction_p_mu}
The probability distribution of the residual magnification $\lambdares=\lambda - \lambdaest$ for sources at redshift $\zS = 1.5$.
Compared are the distribution for $\lambdaest = 0$ (solid line), the distribution $p_{\lambdares}(\lambdares)$ for the optimal estimate \eqref{eq:optimal_magnification_estimate} with $\kappaest$ reconstructed using the shear from an advanced survey with $\ngal=100\,\arcmint^{-2}$ and a filter scale $\thetas = 25\,\arcsect$ (dashed line), and the conditional distribution $p_{\lambdares|\kappaest}(\lambdares|\kappaest)$ for $\kappaest = -0.05$ (dotted line) and $\kappaest=0.1$ (dash-dotted line).
}
\end{figure}

In Fig.~\ref{fig:advanced_reconstruction_p_mu}, the probability distribution of the uncorrected magnification is compared to the distribution of the residual magnification using the optimal estimate \eqref{eq:optimal_magnification_estimate} for correction. The residual distribution is not only narrower, but also less skewed than the distribution of the uncorrected magnification, making it a better candidate for a Gaussian approximation.

In a statistical analysis, one might prefer to use the individual probability distributions $p_{\lambdares|\kappaest}(\lambdaresi|\kappaesti)$ of the residuals $\lambdaresi$ appropriate for the individual convergence estimates $\kappaesti$ instead of a common residual distribution $p_{\lambdares}(\lambdares)$. As Fig.~\ref{fig:advanced_reconstruction_p_mu} shows, the conditional distribution for an estimated convergence $\kappaest=-0.05$ is very narrow with a dispersion that is only 45\% of the dispersion of the distribution of the magnification over the full sky. A standard siren with a convergence estimate that low may thus provide substantially tighter constraints on cosmological parameters than a standard siren without a convergence estimate. On the other hand, a convergence estimate $\kappaest=0.1$ indicates a very broad magnification distribution, which yields a residual dispersion that is twice the dispersion of the full-sky magnification distribution. Knowledge of such a high convergence estimate for a standard siren may help to avoid biases and underestimated errors on its distance due to lensing.

\begin{figure}
\centerline{\includegraphics[width=1\linewidth]{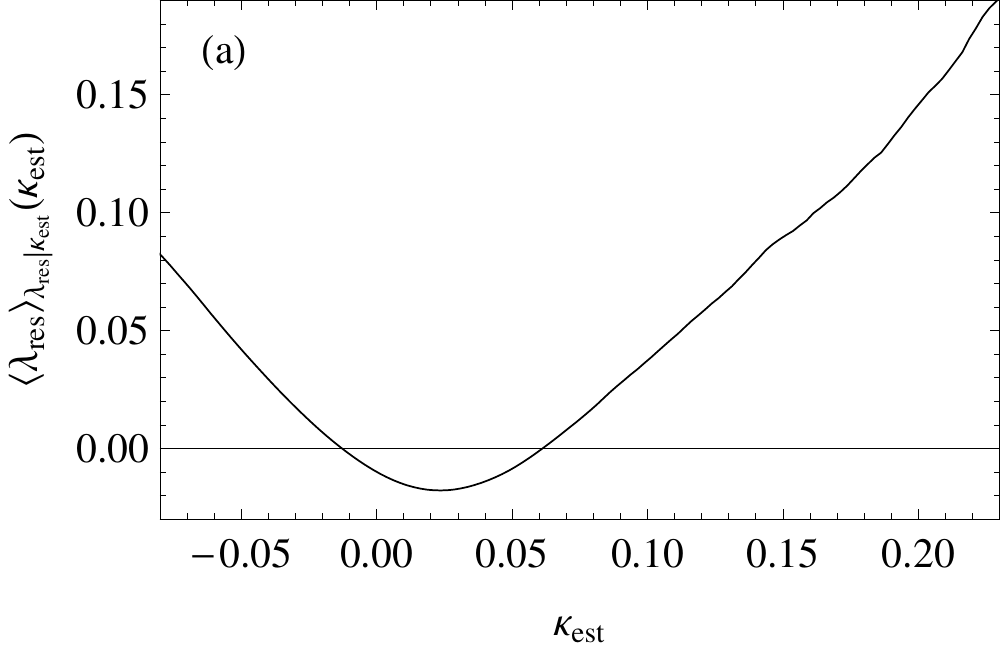}}
\centerline{\includegraphics[width=1\linewidth]{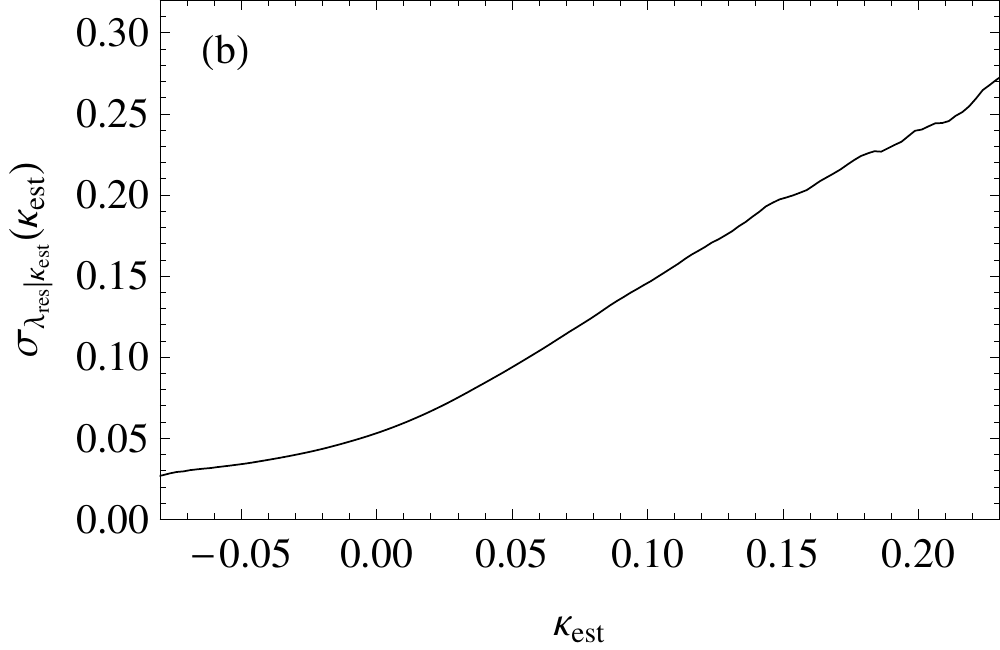}}
\caption{
\label{fig:advanced_reconstruction_residual_conditional_bias_and_dispersion}
Bias $\EV{\lambdares}_{\lambdares|\kappaest}$ and dispersion $\sigma_{\lambdares|\kappaest}$ of the conditional probability distribution $p_{\lambdares|\kappaest}(\lambdares|\kappaest)$ of the residual magnification $\lambdares = \lambda - \lambdaest$ for sources at redshift $\zS = 1.5$ as a function of the estimated convergence $\kappaest$. Shown are the conditional bias and dispersion for the simple estimate \eqref{eq:simple_magnification_estimate} with the convergence $\kappaest$ reconstructed using the shear from an advanced survey with $\ngal=100\,\arcmint^{-2}$ and a filter scale $\thetas = 25\,\arcsect$. Note that the conditional bias of the optimal estimate \eqref{eq:optimal_magnification_estimate} vanishes by construction, while its conditional dispersion is equal to that of the simple estimate.
}
\end{figure}

The dispersion $\sigma_{\lambdares|\kappaest}$ of the conditional probability distribution $p_{\lambdares|\kappaest}(\lambdares|\kappaest)$ as a function of the estimated convergence $\kappaest$ from an advanced galaxy-shear survey is shown in Fig.~\ref{fig:advanced_reconstruction_residual_conditional_bias_and_dispersion}. This dispersion is the same for both the simple estimate \eqref{eq:simple_magnification_estimate} and the optimal estimate \eqref{eq:optimal_magnification_estimate} (see Appendix~\ref{sec:app_magnification_from_convergence}). The simple estimate, however, produces a conditional bias $\EV{\lambdares}_{\lambdares|\kappaest}$, which is absent (by construction) for the optimal estimate.

\begin{figure}
\centerline{\includegraphics[width=1\linewidth]{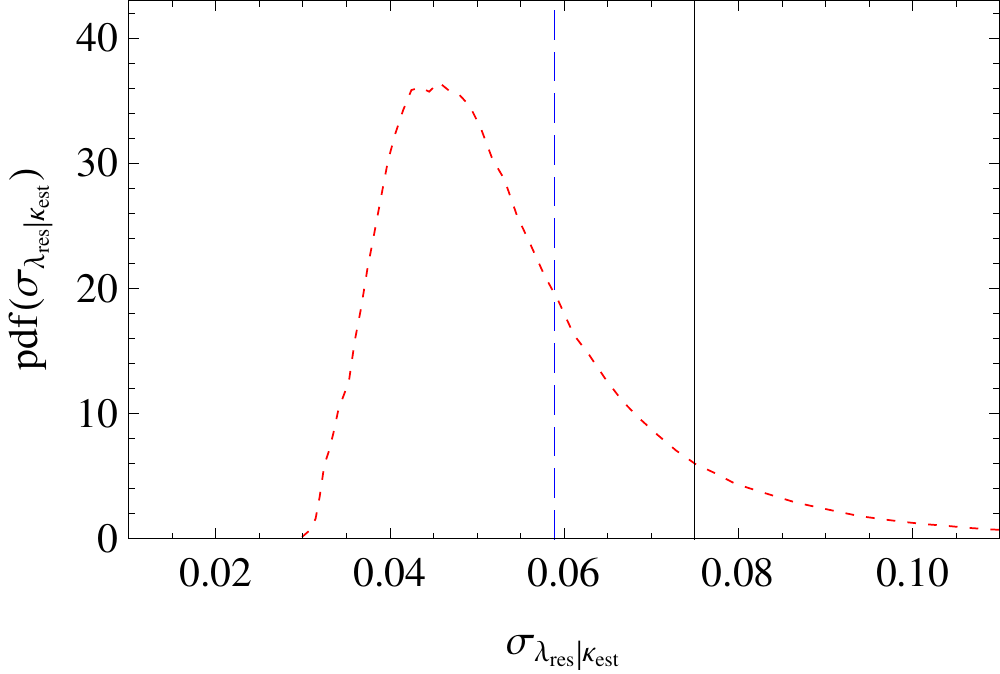}}
\caption{
\label{fig:advanced_reconstruction_residual_conditional_dispersion_pdf}
The probability distribution of the conditional residual dispersion $\sigma_{\lambdares|\kappaest}$ for sources at redshift $\zS = 1.5$ for an advanced survey with shear (dashed line). The vertical solid line indicates the value of the uncorrected magnification dispersion $\sigma_\lambda$, and the vertical dotted line marks the all-sky average $\sigma_{\lambdares}$ of the residual dispersion.
}
\end{figure}

The small residual dispersion in regions with low values of the estimated convergence appears much more encouraging than the all-sky average residual dispersion (compare, e.g., the 45\% residual for $\kappaest=-0.05$ to the all-sky average residual of 78\% a shear survey). The question is, how abundant are regions with small residual dispersion. Figure \ref{fig:advanced_reconstruction_residual_conditional_dispersion_pdf} shows that the distribution of the conditional residual dispersion is very skewed. Values substantially below the all-sky average are common. For example, the median is at $\sigma_{\lambdares|\kappaest} = 0.05$ for an advanced shear survey, which is 10\% below the all-sky average.

\begin{figure}
\centerline{\includegraphics[width=1\linewidth]{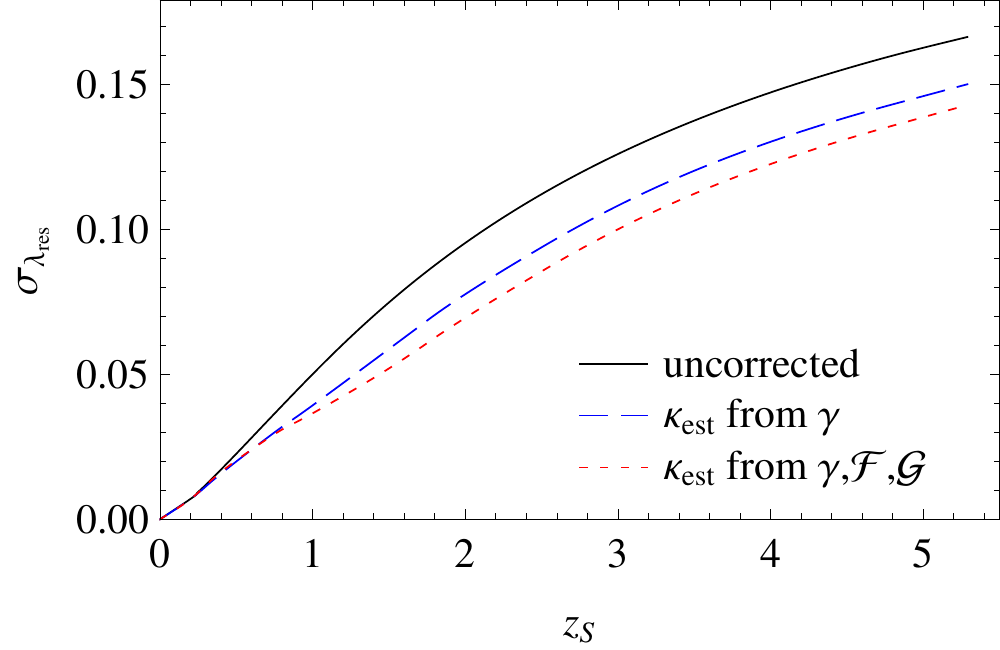}}
\caption{
\label{fig:advanced_reconstruction_residual_dispersion_hi_z}
The dispersion $\sigma_{\lambdares}$ of the residual magnification as a function of the source redshift $\zS$.
The dispersion without correction ($\lambdaest = 0$, solid line) is compared to the residual dispersion with convergence reconstructed from the shear (dashed line) and shear and flexion of an advanced weak-lensing survey with galaxy number density $\ngal=100\,\arcmint^{-2}$ and median redshift $\zmedian = 1.5$.
}
\end{figure}

As Fig.~\ref{fig:advanced_reconstruction_residual_dispersion_hi_z} illustrates, there is little benefit from the magnification correction for standard candles/sirens at redshifts $\zS$ much smaller than the median redshift $\zmedian = 1.5$ of the weak-lensing survey. For $\zS \lesssim 0.5$, a substantial part of the reconstructed convergence is due to matter structures at or above $\zS$. These matter structures shear most galaxy images used for the convergence reconstruction, but do not (de-)magnify the standard candle/siren, and thus merely act as noise in the magnification estimation.

In contrast, standard candles/sirens at redshifts $\zS > \zmedian$ do benefit from delensing. The residual dispersion $\sigma_{\lambdares}$ for sources at $\zS = 3.1$ can be reduced to 85\% of the dispersion $\sigma_{\lambda}$ for the uncorrected magnification, if only shear information is available. If flexion can be used in addition, the dispersion can be further reduced to 80\% of the uncorrected dispersion.

Even for standard candles/sirens at redshift $\zS = 5.3$, there is a noticeable reduction in the residual dispersion. The magnification correction based on shear yields a residual dispersion $\sigma_{\lambdares} = 0.9 \sigma_{\lambda}$ for the best filter scale $\thetas = 25\,\arcsect$. The best correction using shear and flexion reduces the residual dispersion to $85\%$ of the uncorrected dispersion.

\begin{figure}
\centerline{\includegraphics[width=1\linewidth]{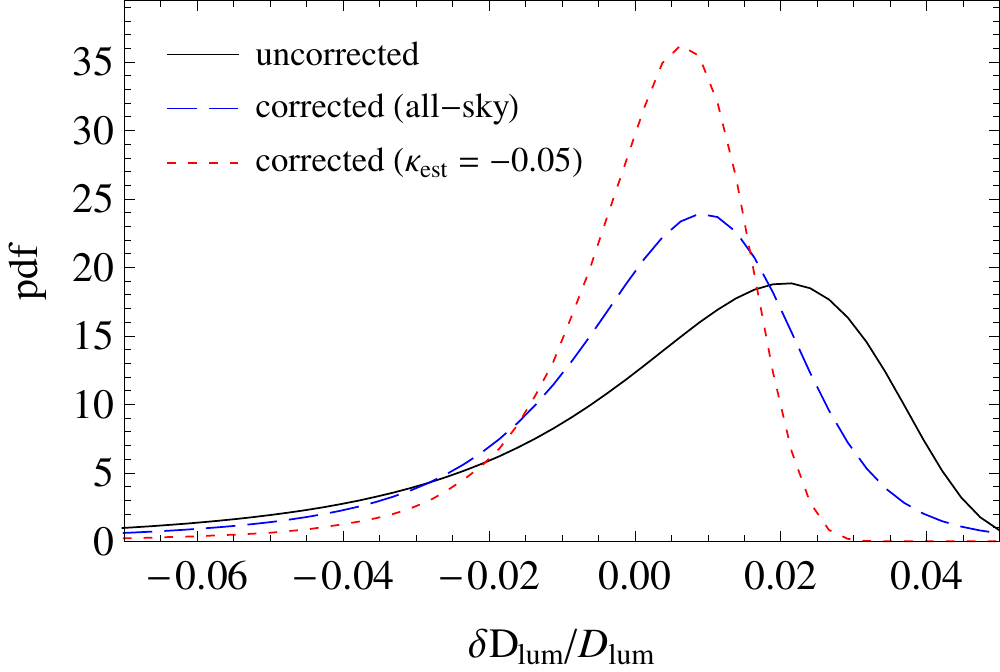}}
\caption{
\label{fig:advanced_reconstruction_p_DeltaDlum}
The probability distribution of the magnification-induced distance error $\delta\Dlum$ relative to the true distance $\Dlum$ for sources at redshift $\zS = 1.5$.
Compared are the distribution without correction (solid line), the all-sky distribution for the corrected distance using the optimal estimate \eqref{eq:optimal_magnification_estimate} with $\kappaest$ reconstructed using the shear from an advanced survey with $\ngal=100\,\arcmint^{-2}$ and a filter scale $\thetas = 25\,\arcsect$ (dashed line), and the distribution for sources in regions with estimated convergence $\kappaest = -0.05$ (dotted line).
}
\end{figure}
\begin{table}
\center
\caption{
\label{tab:advanced_reconstruction_f_small_error}
The fraction of sources at redshift $\zS$ with relative lensing-induced distance error $\varepsilon=|\delta\Dlum/\Dlum|$ below a given threshold  before/after correction with shear data from an advanced survey.
}
\begin{tabular}{r c c c c}
\hline
$ $  & $\varepsilon\leq 0.01$ & $\varepsilon\leq 0.02$ & $\varepsilon\leq 0.05$ & $\varepsilon\leq 0.10$ \\
\hline
$\zS=1.0$ & 0.41/0.60 & 0.80/0.87 & 0.96/0.98 & 0.99/1.00 \\
$\zS=1.5$ & 0.25/0.38 & 0.50/0.69 & 0.92/0.95 & 0.98/0.99 \\
$\zS=2.1$ & 0.18/0.27 & 0.37/0.51 & 0.82/0.89 & 0.96/0.98 \\
$\zS=3.1$ & 0.13/0.18 & 0.27/0.36 & 0.65/0.76 & 0.93/0.95 \\
$\zS=4.2$ & 0.11/0.14 & 0.22/0.28 & 0.54/0.65 & 0.88/0.92 \\
$\zS=5.3$ & 0.10/0.12 & 0.20/0.25 & 0.49/0.58 & 0.83/0.88 \\
\hline
\end{tabular}
\end{table}

The distributions of the lensing-induced distance errors $\delta\Dlum$ are shown in Fig.~\ref{fig:advanced_reconstruction_p_DeltaDlum} for sources at $\zS=1.5$. Before correction, only $50\%$ of the standard candles/sirens  have a relative distance error $\leq2\%$. After correction with shear data from an advanced survey, the fraction rises to $69\%$. In regions with $\kappaest = -0.05$, $90\%$ of the sources have a lensing-induced distance error below $2\%$.
The fraction of sources with absolute lensing-induced distance error below a given threshold before and after correction is listed in Table~\ref{tab:advanced_reconstruction_f_small_error} for several source redshifts.

\subsection{Futuristic reconstruction}
\label{sec:futuristic_reconstruction}

Improvements in observation instruments and techniques might one day permit weak-lensing surveys with very high galaxy number densities. As a futuristic scenario, we consider a very deep high-resolution survey that provides shear and flexion estimates from galaxies with a density $\ngal=500\,\arcmint^{-2}$ and median redshift $\zmedian = 1.8$.

\begin{figure}
\centerline{\includegraphics[width=1\linewidth]{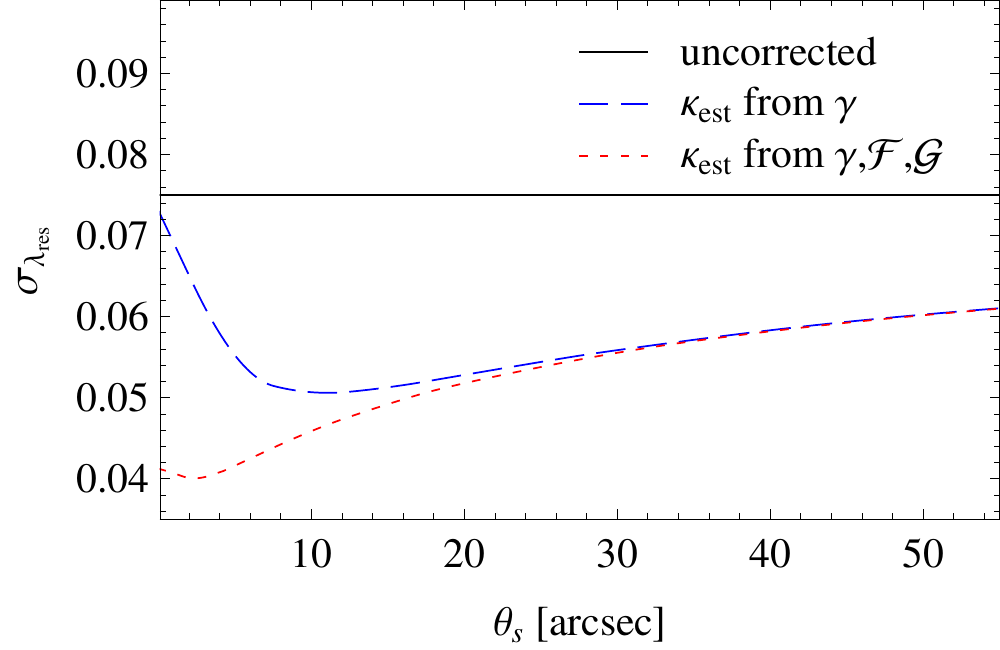}}
\caption{
\label{fig:futuristic_reconstruction_residual_dispersion_and_smoothing}
The dispersion $\sigma_{\lambdares}$ of the residual logarithmic magnification $\lambdares = \lambda - \lambdaest$ of sources at redshift $\zS=1.5$ as a function of the filter scale $\thetas$ used to reconstruct the convergence $\kappaest$ from a futuristic survey with $\ngal=500\,\arcmint^{-2}$. Results are shown for the optimal estimate~\eqref{eq:optimal_magnification_estimate} with convergence reconstructed from shear (dashed line) and from shear and flexion (dotted line). The solid horizontal line indicates the dispersion of $\lambdares=\lambda$ without correction.
}
\end{figure}

As Fig.~\ref{fig:futuristic_reconstruction_residual_dispersion_and_smoothing} shows, the high number density allows one to substantially reduce the residual dispersion $\sigma_{\lambdares}$ for standard candles/sirens at $\zS=1.5$ at small smoothing scales $\thetas$. If only shear data are used, the residual dispersion $\sigma_{\lambdares}$ can be reduced by delensing to 68\% of the uncorrected dispersion $\sigma_\lambda$ at $\thetas=10\,\arcsect$. If shear and flexion data are used, the residual dispersion is $0.04$ at $\thetas=3\,\arcsect$, which is only 54\% of the uncorrected dispersion. To reach a similar statistical uncertainty due to gravitational magnification without delensing, one would need at least a four times higher number of similar standard candles/sirens.

\begin{figure}
\centerline{\includegraphics[width=1\linewidth]{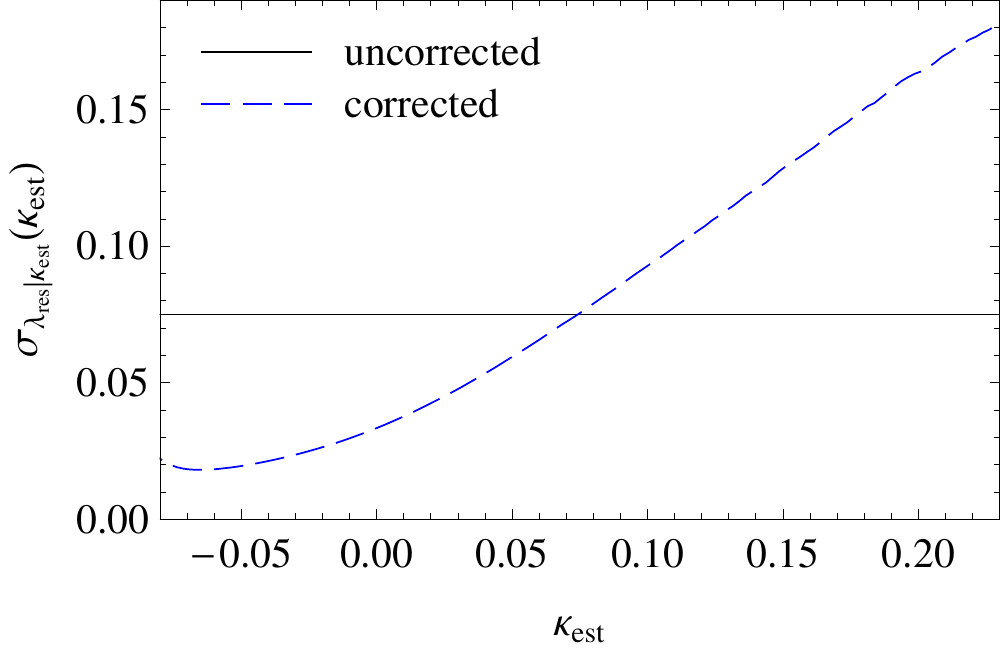}}
\caption{
\label{fig:futuristic_reconstruction_residual_conditional_dispersion}
Dispersion $\sigma_{\lambdares|\kappaest}$ of the conditional probability distribution $p_{\lambdares|\kappaest}(\lambdares|\kappaest)$ 
of the residual magnification $\lambdares=\lambda - \lambdaest$ for sources at redshift $\zS = 1.5$ using the optimal estimate \eqref{eq:optimal_magnification_estimate} and convergence $\kappaest$ reconstructed using the shear and flexion from a futuristic survey with $\ngal=500\,\arcmint^{-2}$ and a filter scale $\thetas = 3\,\arcsect$ (dashed line). The solid line indicates the dispersion of the uncorrected magnification.
}
\end{figure}

For low values of the estimated convergence $\kappaest$, the residual dispersion is even lower, as can be seen in Fig.~\ref{fig:futuristic_reconstruction_residual_conditional_dispersion}. For example, the residual dispersion $\sigma_{\lambdares|\kappaest}$ for $\kappaest=-0.05$ is only 25\% of the unconditional uncorrected dispersion $\sigma_\lambda$.

\begin{figure}
\centerline{\includegraphics[width=1\linewidth]{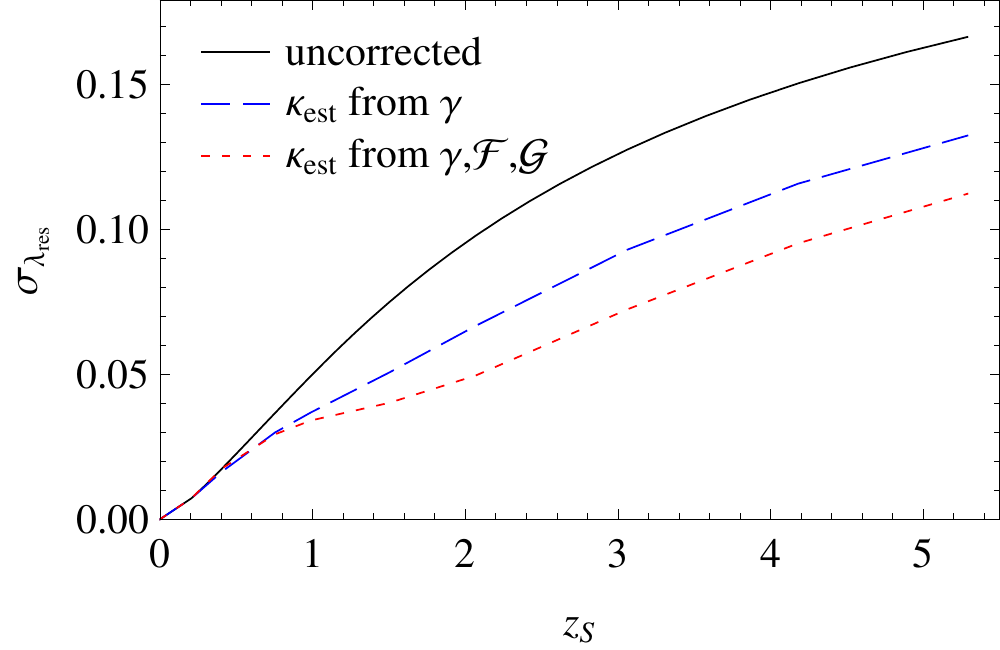}}
\caption{
\label{fig:futuristic_reconstruction_residual_dispersion_hi_z}
The dispersion $\sigma_{\lambdares}$ of the residual magnification as a function of the source redshift $\zS$.
The dispersion without correction ($\lambdaest = 0$, solid line) is compared to the residual dispersion with convergence reconstructed from the shear (dashed line) and shear and flexion of a futuristic weak-lensing survey with galaxy number density $\ngal=500\,\arcmint^{-2}$ and median redshift $\zmedian = 1.8$.
}
\end{figure}

Figure \ref{fig:futuristic_reconstruction_residual_dispersion_hi_z} illustrates that a futuristic survey may also help to substantially reduce the magnifications scatter for high-redshift standard candles/sirens. For example, the residual dispersion for sources at redshift $\zS = 2.1$ can be reduced to $50\%$ of the uncorrected dispersion with a futuristic survey measuring shear and flexion. This is even better than for the sources at redshift $\zS = 1.5$. For $\zS = 5.3$, the residual dispersion can be reduced to $67\%$ of the uncorrected dispersion.

\begin{figure}
\centerline{\includegraphics[width=1\linewidth]{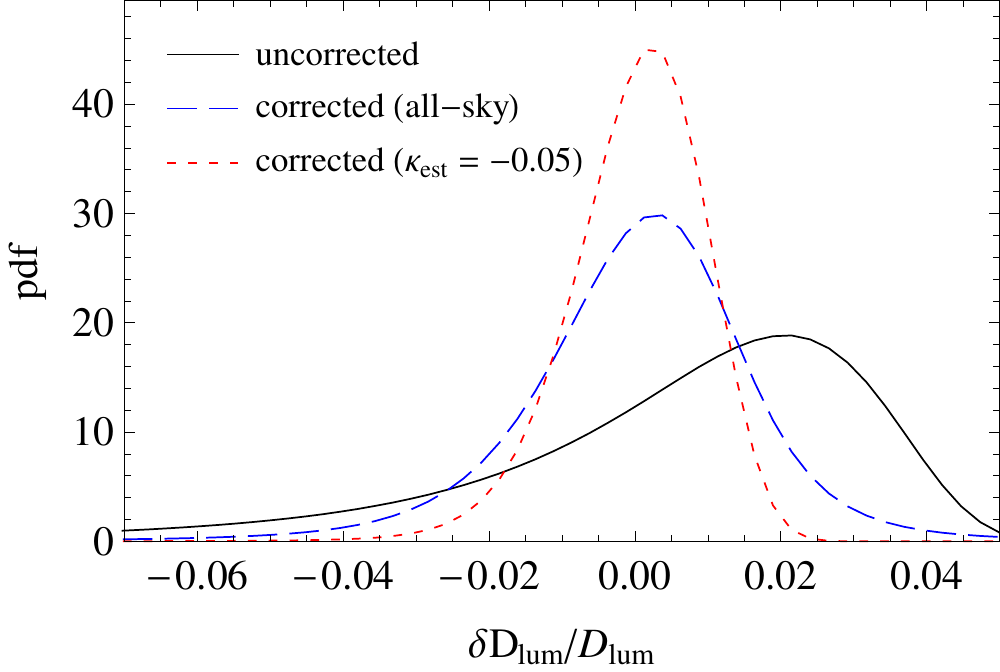}}
\caption{
\label{fig:futuristic_reconstruction_p_DeltaDlum}
The probability distribution of the magnification-induced distance error $\delta\Dlum$ (relative to the true distance $\Dlum$) for sources at redshift $\zS = 1.5$.
Compared are the distribution without correction (solid line), the all-sky distribution for the corrected distance using the optimal estimate \eqref{eq:optimal_magnification_estimate} with $\kappaest$ reconstructed using the shear and flexion from a futuristic survey with $\ngal=500\,\arcmint^{-2}$ and a filter scale $\thetas = 3\,\arcsect$ (dashed line), and the distribution for sources in regions with estimated convergence $\kappaest = -0.05$.
}
\end{figure}

\begin{table}
\center
\caption{
\label{tab:futuristic_reconstruction_f_small_error}
The fraction of sources at redshift $\zS$ with relative lensing-induced distance error $\varepsilon=|\delta\Dlum/\Dlum|$ below a given threshold before/after correction with shear and flexion data from a futuristic survey.
}
\begin{tabular}{r c c c c}
\hline
$ $  & $\varepsilon \leq 0.01$ & $\varepsilon \leq 0.02$ & $\varepsilon \leq 0.05$ & $\varepsilon \leq 0.10$ \\
\hline
$\zS=1.0$ & 0.41/0.64 & 0.80/0.88 & 0.96/0.98 & 0.99/1.00 \\
$\zS=1.5$ & 0.25/0.53 & 0.50/0.82 & 0.92/0.98 & 0.98/1.00 \\
$\zS=2.1$ & 0.18/0.43 & 0.37/0.73 & 0.82/0.97 & 0.96/0.99 \\
$\zS=3.1$ & 0.13/0.29 & 0.27/0.55 & 0.65/0.91 & 0.93/0.98 \\
$\zS=4.2$ & 0.11/0.21 & 0.22/0.41 & 0.54/0.81 & 0.88/0.97 \\
$\zS=5.3$ & 0.10/0.18 & 0.20/0.34 & 0.49/0.73 & 0.83/0.94 \\
\hline
\end{tabular}
\end{table}

Fig.~\ref{fig:futuristic_reconstruction_p_DeltaDlum} shows for sources at $\zS=1.5$, how well a futuristic survey with shear and flexion data can help to mitigate the lensing effects on distance errors. While the fraction of uncorrected standard candles/sirens with relative distance error $\leq2\%$ is only $50\%$, the fraction of corrected standard candles/sirens is $82\%$. If only regions with $\kappaest = -0.05$ are considered, $97\%$ of the sources have residual distance errors $\leq2\%$. Results for several distance error thresholds and source redshifts are listed in Table~\ref{tab:futuristic_reconstruction_f_small_error}.

\subsection{Improving the reconstruction with Wiener filters}
\label{sec:Wiener_filters}

Using a Gaussian filter to smooth the reconstructed convergence [cf. Eq.~\eqref{eq:Gauss_filter}] might not be the best way to exploit the spatial correlation in the convergence field. With a few assumptions, choosing the optimal filter for the reconstructed convergence becomes a text-book problem on Wiener filters \citep[][]{Wiener_book}.

Consider the unfiltered convergence $\kapparaw$ reconstructed from galaxy shear or shear and flexion as a raw estimate for the convergence $\kappa(\zS)$ to a standard candle/siren at redshift $\zS$. Let $P_{\kapparaw}(\ell)$ denote the power spectrum of $\kapparaw$, and $P_{\kappa(\zS),\kapparaw}(\ell)$ denote the cross power spectrum of $\kappa(\zS)$ and $\kapparaw$. Then
\begin{equation}
\label{eq:Wiener_filter}
  \hat{W}(\ell) = \frac{P_{\kappa(\zS),\kapparaw}\bigl(\ell)}{P_{\kapparaw}\bigl(\ell)}
\end{equation}
provides a Wiener filter in Fourier space, which minimises the mean square difference between the `true' convergence $\kappa(\zS)$ and estimated convergence $\kappaest(\vect{\theta})$, given in Fourier space by
\begin{equation}
\label{eq:kappa_Wiener_filter}
  \hatkappaest(\vect{\ell}) = \hat{W}\bigl(|\vect{\ell}|\bigr) \hatkapparaw(\vect{\ell})
.
\end{equation}

\begin{figure}
\centerline{\includegraphics[width=1\linewidth]{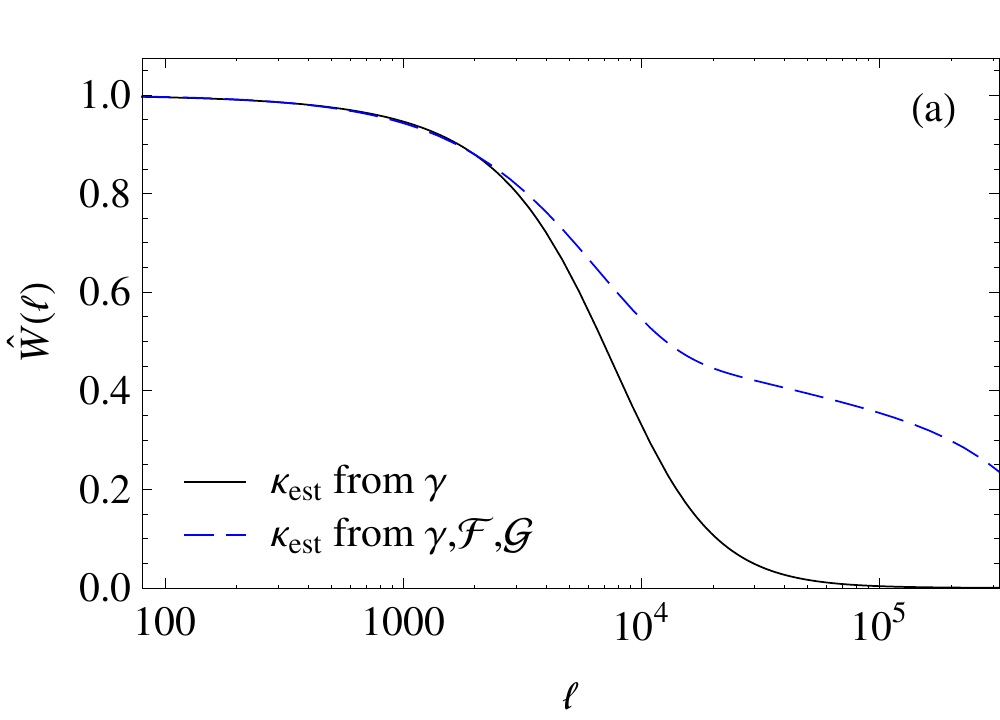}}
\centerline{\includegraphics[width=1\linewidth]{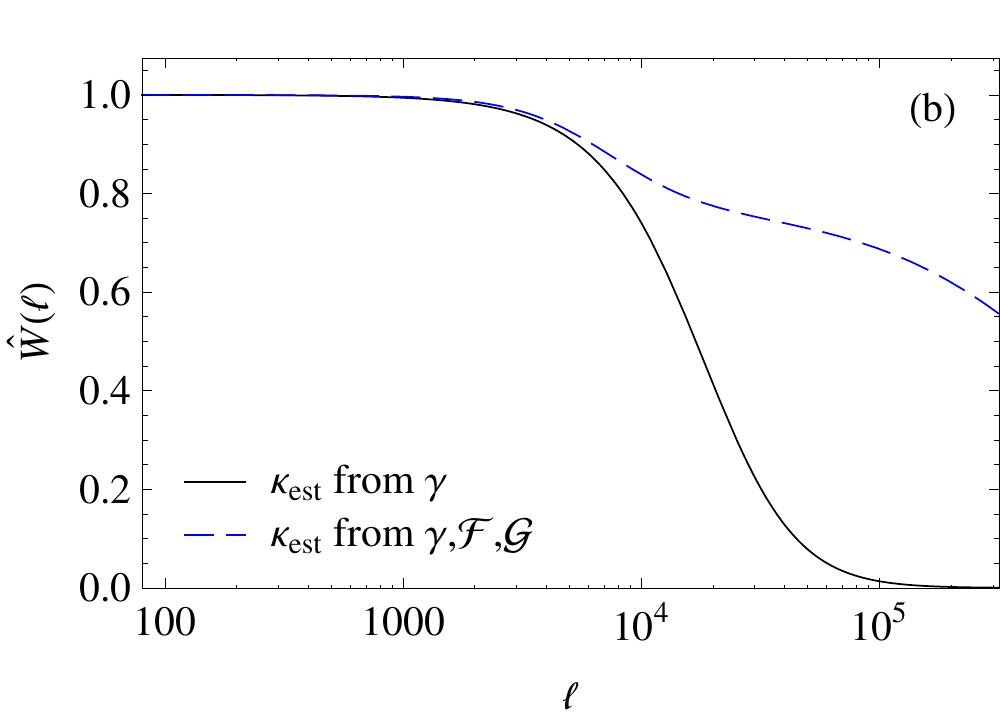}}
\caption{
\label{fig:Wiener_filter}
Wiener filters $\hat{W}(\ell)$ in Fourier space for standard candles/sirens at redshift $\zS=1.5$ and convergence reconstructed from the shear (solid line) or the shear and flexion (dashed line) from an advanced survey with galaxy density $\ngal=100\,\arcmint^{-2}$ and median redshift $\zmedian = 1.5$ (a) and a futuristic survey with  $\ngal=500\,\arcmint^{-2}$ and $\zmedian = 1.8$ (b).
}
\end{figure}

The Wiener filter \eqref{eq:Wiener_filter} can be constructed by measuring the power spectra in the simulated fields and approximating their ratios with a suitable function (e.g. a spline). The filters for standard candles/sirens at redshift $\zS=1.5$ are shown in Fig.~\ref{fig:Wiener_filter} for the advanced and the futuristic survey.

Using Wiener filters for the estimated convergence improves the dispersion in the residual magnification only marginally compared to the best Gaussian smoothing. For an advanced survey with $\ngal=100\,\arcmint^{-2}$ with shear data, the Wiener filter yields a residual dispersion $\sigma_{\lambdares} = 78\%$ of the uncorrected dispersion $\sigma_{\lambda}$, whereas the Gaussian filter with $\thetas=25\,\arcsect$ yields a residual dispersion $\sigma_{\lambdares} = 79\%$. For a futuristic survey with shear and flexion, the residual dispersion is $53\%$ of uncorrected dispersion for the Wiener filter, which is to be compared to $54\%$ for the best Gaussian filter.

\subsection{Improving the reconstruction with individual galaxy redshifts}
\label{sec:redshift_weighting}

The image distortion for a galaxy at redshift $z$, and hence the information in the galaxy image about the magnification at the redshift $\zS$ of a standard candle/siren is not uniform in redshift $z$. In a full tomographic analysis, the variation of the lensing signal with redshift is used to perform a three-dimensional reconstruction of the matter structures in the light cone \citep[e.g.][]{MasseyEtal2007_3D_WL}. The reconstructed three-dimensional matter distribution can then be used to estimate the magnification for a given sky position and redshift. Here, we take a simpler approach, which does not try to locate the actual matter structures causing the lensing in redshift, but only exploits the statistical relation between lensing quantities at different redshifts. 

\begin{figure}
\centerline{\includegraphics[width=1\linewidth]{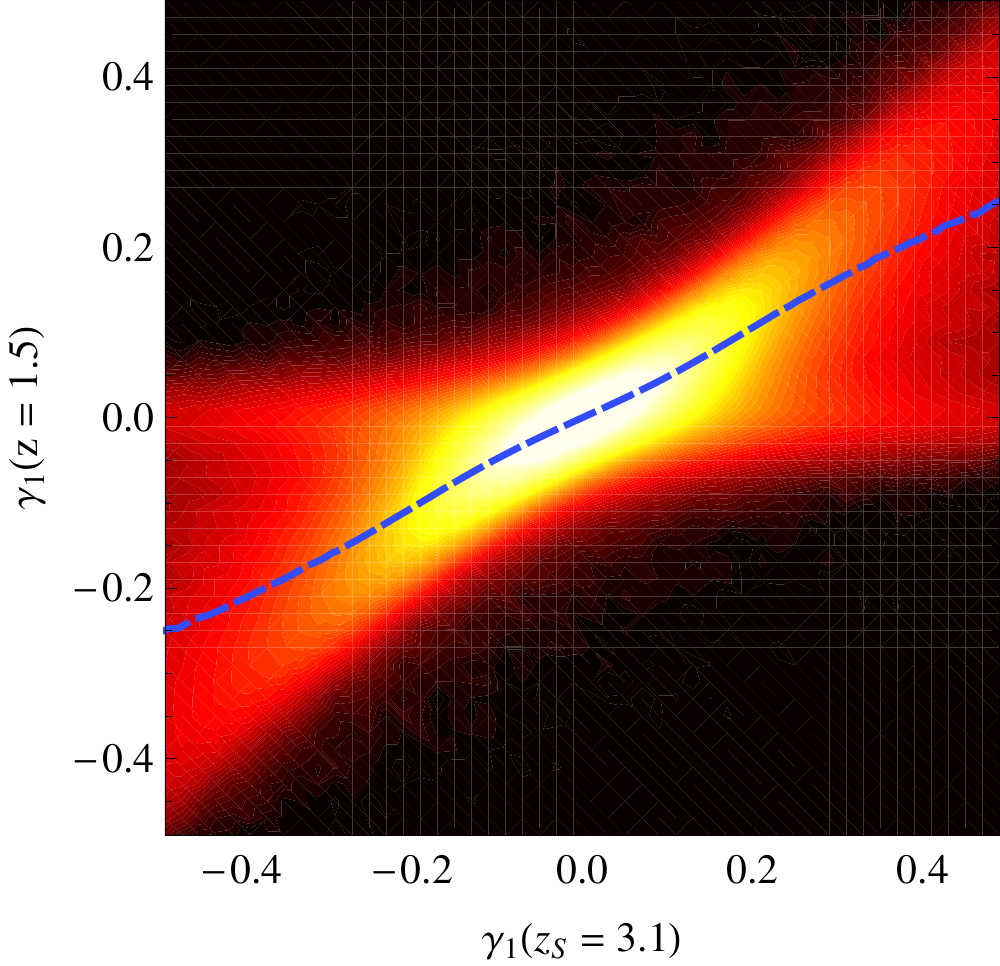}}
\caption{
\label{fig:redshift_weighting_p_gamma_gamma}
The joint distribution of the shear component $\gamma_1$ at redshift $z = 1.8$ and redshift $\zS = 3.1$ at the same sky position (lighter areas correspond to higher probability densities on a logarithmic scale). The dashed line marks the mean shear $\EV{\gamma_1(z)}_{\gamma_1(z)|\gamma_1(\zS)}$ as a function of $\gamma_1(\zS)$. Note that the joint distribution for the second shear components $\gamma_2(z)$ and $\gamma_2(\zS)$ is the same as for the first shear components.
} 
\end{figure}
\begin{figure}
\centerline{\includegraphics[width=1\linewidth]{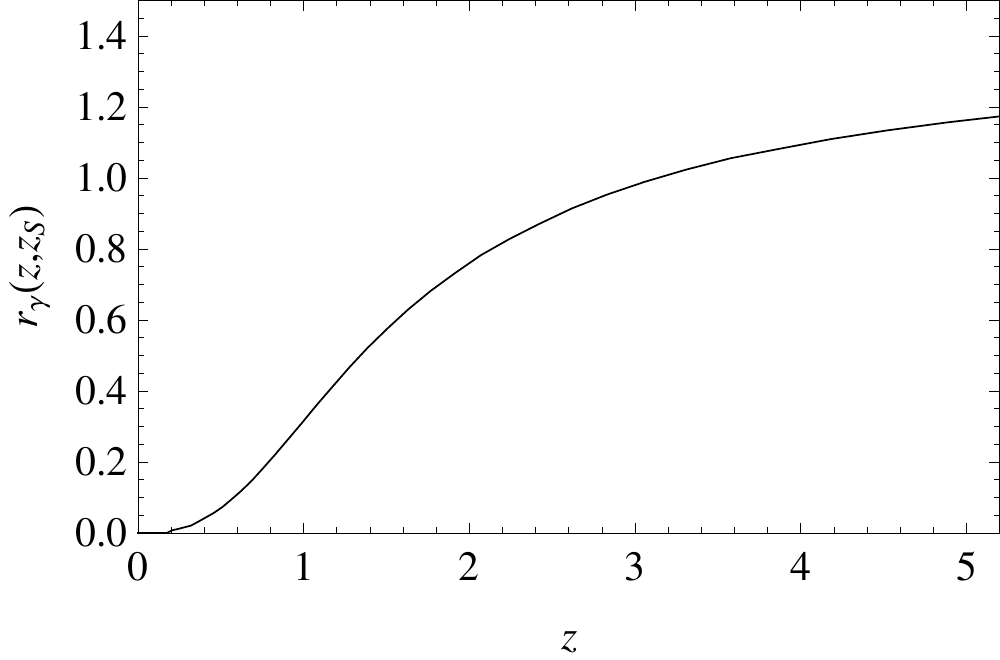}}
\caption{
\label{fig:redshift_weighting_gamma_ratio}
The ratio $r_{\gamma}(z,\zS)$ between the mean shear at redshift $z$ and the shear at redshift $\zS=3.1$. The ratio has been calculated by fitting the relation \eqref{eq:mean_gamma_z_of_gamma_zS_linear_approx} to the shear data from the ray-tracing simulations.
}
\end{figure}

We start from the same assumptions used for the simulated the weak-lensing surveys (see Section \ref{sec:mock_catalogs}): The image shape of a galaxy at sky position $\thetagal$ and redshift $\zgal$ provides noisy but unbiased estimates $\gammagal$, $\Fgal$, and $\Ggal$ of the shear $\gamma(\thetagal,\zgal)$ and flexion $\F(\thetagal,\zgal)$ and $\G(\thetagal,\zgal)$ to the galaxy redshift $\zgal$. The error in these estimates is simply the shape noise with vanishing mean.
Considered as estimates for the shear and flexion $\gamma(\thetagal,\zS)$, $\F(\thetagal,\zS)$, and $\G(\thetagal,\zS)$ at the redshift $\zS$ of a standard candle/siren, $\gammagal$, $\Fgal$, and $\Ggal$ acquire an additional error, namely the deviation between the shear and flexion at different redshifts. In the following discussion, this deviation is split into a systematic (i.e. zero-scatter) and a statistical (i.e. zero-mean) component.

The first step towards an improved reconstruction aims at eliminating the systematic deviation. As Fig.~\ref{fig:redshift_weighting_p_gamma_gamma} illustrates, the shear values for two different redshifts along the same line of sight are strongly correlated, but on average, the shear to the lower redshift is smaller in magnitude. Figure~\ref{fig:redshift_weighting_p_gamma_gamma} also shows that the mean $\EV{\gamma(z)}_{\gamma(z)|\gamma(\zS)}$ of the shear $\gamma(z)$ to redshift $z$ for given shear $\gamma(\zS)$ to redshift $\zS$ along the same line of sight is well approximated by the linear relation
\begin{subequations}
\label{eq:conditional_mean_linear_approx}
\begin{align}
\label{eq:mean_gamma_z_of_gamma_zS_linear_approx}
	\EV{\gamma(z)}_{\gamma(z)|\gamma(\zS)} &= r_{\gamma}(z,\zS) \gamma(\zS)
\intertext{with a redshift-dependent proportionality factor $r_{\gamma}(z,\zS)$ (see Fig.~\ref{fig:redshift_weighting_gamma_ratio}). Moreover,}
\label{eq:mean_F_z_of_F_zS_linear_approx}
	\EV{\F(z)}_{\F(z)|\F(\zS)} &= r_{\F}(z,\zS)\F(\zS)
	\text{ and}\\
\label{eq:mean_G_z_of_G_zS_linear_approx}
	\EV{\G(z)}_{\G(z)|\G(\zS)} &= r_{\G}(z,\zS)\G(\zS)
\end{align}
\end{subequations}
with factors $r_{\F}(z,\zS)$ and $r_{\G}(z,\zS)$ very similar to $r_{\gamma}(z,\zS)$. Thus, one can construct unbiased estimates $\gammagal(\zS)$, $\Fgal(\zS)$, and $\Ggal(\zS)$ of the shear and flexion at redshift $\zS$ by:
\begin{subequations}
\label{eq:scaled_estimates}
\begin{align}
\label{eq:scaled_shear_estimate}
  \gammagal(\zS) &= \bigl[r_{\gamma}(\zgal,\zS)\bigr]^{-1} \gammagal
  , \\
  \Fgal    (\zS) &= \bigl[r_{\F}    (\zgal,\zS)\bigr]^{-1} \Fgal
  \text{, and}\\
  \Ggal    (\zS) &= \bigl[r_{\G}    (\zgal,\zS)\bigr]^{-1} \Ggal
  .
\end{align}
\end{subequations}

In a second step, the statistical weights of the galaxies in the reconstruction are extended to take into account the redshift dependence of the error in the scaled estimates \eqref{eq:scaled_estimates}. The error in these estimates has two components: the shape-noise and the scatter around the mean relation between the lensing quantities at different redshifts. The shape noise in the scaled estimate is simply the shape noise in the unscaled estimate (de-)amplified by the proportionality factor $r_{\gamma}(z,\zS)$, $r_{\F}(z,\zS)$, or $r_{\G}(z,\zS)$, and has no spatial correlation prior to smoothing. The scatter around the mean relation between shear and flexion at different redshifts has a quite different redshift dependence (e.g. it vanishes when $\zgal=\zS$), and has a finite spatial correlation inherited from the intrinsic spatial correlation of the shear and flexion at a single redshift. As a consequence, the optimal redshift weights depend on the galaxy number density and the spatial smoothing. However, an examination of the optimal weights (which can be derived in a lengthy calculation) reveals that the contribution from the scatter around the mean relation is much smaller than the shape-noise contribution for the smoothing scales and number densities of interest (at least for $z\lesssim\zS$). Thus one can approximate the redshift weights by:
\begin{subequations}
\label{eq:redshift_weights}
\begin{align}
\begin{split}
  w_{z,\gamma}(z, \zS) &= r_\gamma(z, \zS)^{2}
,\\
  w_{z,\F}   (z, \zS) &= r_\F    (z, \zS)^{2}
\text{, and}\\
  w_{z,\G}   (z, \zS) &= r_\G    (z, \zS)^{2}
.
\end{split}
\end{align}
\end{subequations}

Optimal use of the scaled estimates \eqref{eq:scaled_estimates} and redshift-dependent weights \eqref{eq:redshift_weights} requires that the information about the error of the scaled estimates is still available for the spatial smoothing of the resulting fields. Thus, the reconstruction algorithm discussed in Section ~\ref{sec:reconstruction} is modified by merging the spatial-filtering step with the step where the galaxy estimates are projected onto a regular mesh. Since the smoothing is now computed in real space, the algorithm is much slower than the one described Section~\ref{sec:reconstruction}, in particular for larger smoothing scales.

\begin{figure}
\centerline{\includegraphics[width=1\linewidth]{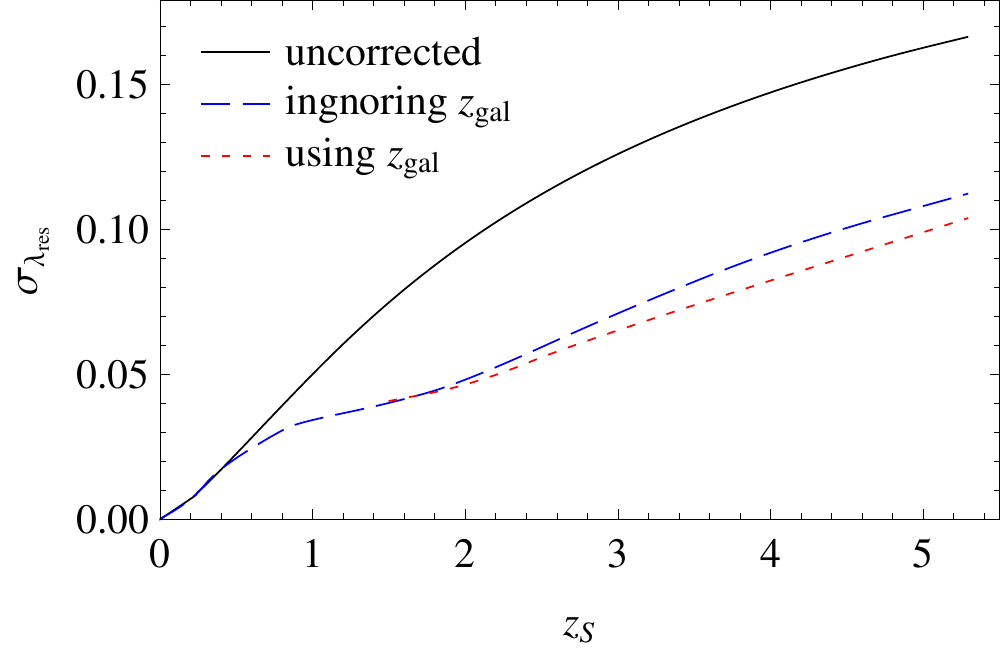}}
\caption{
\label{fig:redshift_weighting_residual_dispersion_of_z}
The dispersion $\sigma_{\lambdares}$ of the residual magnification as a function of the source redshift $\zS$.
The dispersion without correction ($\lambdaest = 0$, solid line) is compared to the residual dispersion with convergence reconstructed from the shear and flexion of a futuristic weak-lensing survey with galaxy number density $\ngal=500\,\arcmint^{-2}$ and median redshift $\zmedian = 1.8$, when ignoring (dashed line) or using the individual galaxy redshifts (dotted line).
}
\end{figure}

For standard candles/sirens at redshift $\zS = 1.5$, there is no improvement in the reconstruction from an advanced or futuristic survey (shown in Fig.~\ref{fig:redshift_weighting_residual_dispersion_of_z}). This is different for higher redshifts. For $\zS = 3.1$, the reconstruction from the shear and flexion of a futuristic survey changes from $57\%$ to $52\%$ of the uncorrected dispersion if redshift weights are used. For $\zS = 5.3$, redshift weighting improves the residual dispersion from $67\%$ to $62\%$.

\section{Summary and Discussion}
\label{sec:summary}

Unless corrected for, gravitational lensing induces significant errors in the measured distances of high-redshift standard candles and standard sirens.
In this work, we have used numerical simulations to investigate how much these distance errors can be reduced with weak-lensing reconstruction. The method considered comprises (i) reconstructing the convergence towards a standard candle/siren from the shear and flexion measured in a galaxy lensing survey; (ii) estimating the magnification from the reconstructed convergence; and (iii) correcting the observed signal of the standard candle/siren using the inferred magnification.

By measuring the relation between the reconstructed convergence and the magnification, we have constructed an optimised magnification estimate that is unbiased and minimises the residual magnification errors. Furthermore, we have studied the optimal smoothing scale for the weak-lensing reconstruction under various survey conditions.

For an advanced shear survey with $\ngal = 100\,\arcmint^{-2}$, we find that the lensing-induced distance errors for standard candles/sirens at redshifts $\zS\approx1.5$ are reduced on average by $20\%$. This confirms earlier findings by \citet{DalalEtal2003} that reconstructions based on such surveys do not significantly reduce the lensing errors on average. However, in regions with low estimated convergence $\kappaest$, the residual magnification uncertainty is substantially smaller than in regions of high convergence (e.g. only $40\%$ of the uncorrected error for $\kappaest = -0.05$ compared to $200\%$ for $\kappaest = 0.1$). Thus a weak-lensing survey is useful to identify standard candles/sirens in regions with low (high) convergence and small (large) magnification errors, which can then be given larger (smaller) weights in a statistical analysis.

As already pointed out by \citet{ShapiroEtal2010}, the correction from weak-lensing reconstruction can be improved considerably by including flexion measurements and greatly increasing the galaxy number density. For example, a futuristic shear and flexion survey with $\ngal = 500\,\arcmint^{-2}$ yields error reductions of 50\% for standard candles/sirens at $\zS \sim1.5$, and 35\% for sources at $\zS\sim5$. In low-convergence regions, the errors can be reduced by up to $75\%$. For example, the residual distance errors are below 2\% for over 97\% of the standard candles/sirens in regions with $\kappaest=-0.05$. Such lensing-corrected standard candles/sirens would constitute very competitive distance indicators (but one should keep in mind that these numbers are based on very optimistic assumptions about the properties of future surveys).

Our simulation provides detailed statistical information about the magnification-induced distance errors before and after correction, in particular their  probability distributions as a function of the estimated convergence. The distributions are valuable for Bayesian parameter estimation and model selection. In a forthcoming paper \citetext{Gair, King, \& Hilbert, in prep.}, we will discuss the resulting implications for the use of high-redshift standard sirens as cosmological probes.

In this paper, we have considered several ways to improve the accuracy of the magnification correction scheme. Wiener filters constructed from the measured (cross-)power spectra of the true and reconstructed convergence are expected to provide the optimal filtering of the noisy convergence maps. However, in our simulations, they perform only marginally better than Gaussian filters with suitable filter scale.

Furthermore, we have investigated a simple method employing redshift-dependent weights for the galaxies to improve the weak-lensing reconstruction. While standard candles/sirens at redshifts $\zS\leq2$ show no benefit from this method, magnification errors for higher redshifts are reduced by an additional 5\%. Further improvement might come from a tomographic reconstruction \citep[][]{SimonTaylorHartlap2009}.

In contrast to correction schemes based on modelling the foreground matter structures from the observed light \citep[e.g.][]{GunnarssonEtal2006}, the method discussed here has the advantage that it does not require any assumptions about the relation between visible and dark matter. On the other hand, the light distribution contains valuable information about the matter distribution and the magnification. A better magnification correction might be obtained by inferring the foreground matter distribution from both the observed properties of galaxies along the line of sight and the shear and flexion information from distant galaxies. Additional information about the matter distribution at redshifts beyond those probed by conventional galaxy weak lensing and flexion could be gathered from observations of lensing of the cosmic microwave background \citep[][]{LewisChallinor2006} or high-redshift 21-cm radiation \citep[][]{HilbertMetcalfWhite2007}.

In our simulations, we have fully taken into account many aspects of weak-lensing galaxy surveys that impact the accuracy of the correction. These include, e.g., the non-Gaussian nature of the convergence field, the randomness and discreteness of the galaxy redshifts and image positions, and the complex relation between the reconstructed convergence and the magnification. Thus, our studies provide more realistic predictions about the accuracy of the correction than earlier studies relying on convergence power spectra.

However, our approach neglects several complications that affect the correction. Future studies should include the effects of observing reduced shear and flexion. The impact of source clustering and intrinsic alignment on the weak-lensing reconstruction must be investigated.

Our simulations do not include the effects of lensing by structures on scales below the resolution of the MS (i.e. dark-matter structures on scales $<5h^{-1}\,\kpc$ or luminous structures with masses $<10^9h^{-1}\,\Msolar$). Such structures could affect the magnification of SNe, GRBs, and SMBs, but it would be difficult to recover them in a weak-lensing mass reconstruction. The resulting increase in the uncorrected and residual magnification error should be investigated, possibly with higher-resolution simulations. One should also consider using the observed scatter in samples of standard candles/sirens to constrain the small-scale matter fluctuations missed by the simulations. 

Finally, the cosmology dependence of the correction should be explored. For example, a higher/lower normalisation $\sigma_8$ increases/decreases both the uncorrected and the residual magnification dispersion. Moreover, errors in the assumed cosmology lead to a suboptimal magnification estimator and thus increase the residual magnification error and might also introduce a bias.  The resulting degradation should be quantified in future studies. One should also investigate how well one could detect a possible bias in the magnification estimates by looking at differences between the expected and observed distributions of the convergence and magnification estimates.

\section*{Acknowledgments}
We thank Xinzhong Er, Jan Hartlap, Antony Lewis, Peter Schneider, and Simon White for helpful discussions. This work was supported by the DFG within the Priority Programme 1177 under the projects SCHN 342/6 and WH 6/3 (SH), and by the Royal Society (JRG and LJK).


\appendix

\section{More about magnification estimates from noisy convergence maps}
\label{sec:app_magnification_from_convergence}

Define for any two random variables $x$ and $y$ with joint probability density function (pdf) $p_{x,y}(x,y)$ and conditional pdf $p_{x|y}(x|y)$, and any function $f(x,y)$ as short-hand notation
\begin{equation}
	\EV{f} = \EV{f}_{x,y} = \iint{}\!\!\diff[]{x}\,\diff[]{y}\,p_{x,y}(x,y) f(x,y)
\end{equation}
for the unconditional mean of $f$, and 
\begin{equation}
\EV{f}_{|y}(y) = \EV{f}_{x|y}(y) = \int{}\!\!\diff[]{x}\,p_{x|y}(x|y) f(x,y)
\end{equation}
for the conditional mean of $f$ for given $y$. Furthermore, define
\begin{equation}
	\sigma^2_{f} =  \iint{}\!\!\diff[]{x}\,\diff[]{y}\,p_{x,y}(x,y) \bigl[ f(x,y) - \EV{f} \bigr]^2
\end{equation}
for the full variance of $f$, and 
\begin{equation}
	\sigma^2_{f|y}(y)=\int{}\!\!\diff[]{x}\,p_{x|y}(x|y) \bigl[ f(x,y) - \EV{f}_{x|y} \bigr]^2
\end{equation}
for the variance of $f$ restricted to a fixed $y$.

Consider an ensemble of lines of sight (l.o.s.), where each l.o.s. is characterised by its (logarithmic) magnification $\lambda$ and its estimated convergence $\kappaest$. Assume that the joint pdf $p_{\lambda,\kappaest}(\lambda,\kappaest)$ of $\lambda$ and $\kappaest$ is known for the ensemble, which implies that the marginal and conditional distributions of $\lambda$ and $\kappaest$ are known, too. Furthermore, assume the convergence $\kappaest$ known for each l.o.s.

Consider a magnification estimate $\lambdaest = \lambdaest(\kappaest)$ that is based on the estimated convergence $\kappaest$. Consider the (unknown) residual magnification 
\begin{equation}
\lambdares = \lambda - \lambdaest(\kappaest)
\end{equation}
obtained after correcting the (unknown) magnification $\lambda$ of a l.o.s. by the magnification estimate $\lambdaest(\kappaest)$ obtained from its (known) convergence estimate $\kappaest$. The conditional bias $\EV{\lambdares}_{\lambda|\kappaest}$, i.e. the mean residual for a fixed estimated convergence $\kappaest$, is then given by:
\begin{equation}
 \EV{\lambdares}_{\lambda|\kappaest}(\kappaest) = \EV{\lambda}_{\lambda|\kappaest}(\kappaest) - \lambdaest(\kappaest).
\end{equation}
Moreover, the conditional distribution $p_{\lambdares|\kappaest}(\lambdares|\kappaest)$ of the residual $\lambdares$ for a given convergence estimate $\kappaest$ is then simply a shifted version of the conditional pdf $p_{\lambda|\kappaest}(\lambda|\kappaest)$ of the true magnification $\lambda$:
\begin{equation}
\label{eq:cond_res_pdf_from_cond_pdf}
	p_{\lambdares|\kappaest}(\lambdares|\kappaest) = p_{\lambda|\kappaest}\bigl(\lambdares + \lambdaest(\kappaest)|\kappaest\bigr).
\end{equation}
This implies that the conditional variance $\sigma^2_{\lambdares|\kappaest}$ of the residual equals the conditional variance $\sigma^2_{\lambda|\kappaest}$ of the true magnification for any magnification estimate $\lambdaest(\kappaest)$.

The problem of finding the optimal magnification estimate $\lambdaest(\kappaest)$ can be approached with basic calculus of variations. First, one has to define `optimal', e.g. by demanding that an optimal estimate (i) is unbiased, i.e. $\bEV{\lambdaest}_{\lambda|\kappaest} = \bEV{\lambda}_{\lambda|\kappaest}$, and (ii) minimises the residual variance\footnote{
From any biased estimate $\tlambdaest(\kappaest)$, one can construct an unbiased estimate that yields the same residual variance by: 
$\lambdaest(\kappaest) = \tlambdaest(\kappaest) - \bEV{\tlambdaest}_{\lambda|\kappaest}  + \bEV{\lambda}_{\lambda|\kappaest}$. 
Thus, one can restrict the discussion to unbiased estimates without missing estimates with minimal residual variance.
}
\begin{equation}
\begin{split}
\sigma^2_{\lambdares} &= 
  \!\!\iint\!\!\diff[]{\lambda}\,\diff[]{\kappaest}\,p_{\lambda,\kappaest}(\lambda,\kappaest)
  \bigl[ \lambda - \lambdaest(\kappaest) \bigr]^2
\\&=
  \!\!\int\!\!\diff[]{\kappaest}\,p_{\kappaest}(\kappaest)
  \\&\quad\times
  \!\!\int\!\!\diff[]{\lambda}\,p_{\lambda|\kappaest}(\lambda|\kappaest)
  \bigl[ \lambda - \lambdaest(\kappaest) \bigr]^2
 .  
\end{split}  
\end{equation}
Then, 
\begin{equation}
\begin{split}
0 &= 
\frac{\partial}{\partial \lambdaest }
  \!\!\int\!\!\diff[]{\lambda}\,p_{\lambda|\kappaest}(\lambda|\kappaest)
  \bigl[ \lambda - \lambdaest(\kappaest) \bigr]^2
\\&=
  \!\!\int\!\!\diff[]{\lambda}\,p_{\lambda|\kappaest}(\lambda|\kappaest)
  \bigl[ \lambda - \lambdaest(\kappaest) \bigr]
 \\&=
  \EV{\lambda}_{\lambda|\kappaest}(\kappaest) - \lambdaest(\kappaest)
\end{split}  
\end{equation}
for the optimal estimate. Thus, the optimal estimate reads:
\begin{equation}
 \lambdaest(\kappaest) = \EV{\lambda}_{\lambda|\kappaest}(\kappaest)
 .
\end{equation}



\end{document}